\documentclass[10pt,aps,prd,twocolumn,superscriptaddress,preprintnumbers,showpacs, floatfix, nofootinbib]{revtex4-1}
\usepackage{amsmath,latexsym,
amsfonts,amssymb,graphicx,psfrag,dsfont,slashed,bigints,units,bm}
\usepackage[colorlinks=true,citecolor=blue]{hyperref}
\usepackage[normalem]{ulem}
\usepackage[dvipsnames]{xcolor}
\usepackage[caption=false]{subfig}
\newcommand{\de}{\partial}

\newcommand{\be}{\begin{equation}}
\newcommand{\ee}{\end{equation}}
\newcommand{\ba}{\begin{eqnarray}}
\newcommand{\ea}{\end{eqnarray}}

\newcommand{\bb}{\mathrm{B}}

\DeclareMathOperator*{\argmin}{arg\,min}

\newcommand{\SECTION}[1]{{\smallskip\medskip\noindent\textbf{#1.}}}

\begin{document}

\title{Superfluid dark stars}
\author{Lorenzo Cipriani}
\email[]{lorenzo.cipriani@graduate.univaq.it}
\affiliation{Dipartimento di Scienze Fisiche e Chimiche,
Università dell’Aquila, via Vetoio, I-67100, L’Aquila, Italy}
\affiliation{INFN Laboratori Nazionali del Gran Sasso, Via G. Acitelli
22, 67100 Assergi (AQ), Italy}
\author{Massimo Mannarelli}
\email[]{massimo.mannarelli@lngs.infn.it}
\affiliation{INFN Laboratori Nazionali del Gran Sasso, Via G. Acitelli
22, 67100 Assergi (AQ), Italy}
\author{Fabrizio Nesti}
\email[]{fabrizio.nesti@aquila.infn.it}
\affiliation{Dipartimento di Scienze Fisiche e Chimiche,
Università dell’Aquila, via Vetoio, I-67100, L’Aquila, Italy}
\affiliation{INFN Laboratori Nazionali del Gran Sasso, Via G. Acitelli
22, 67100 Assergi (AQ), Italy}
\author{Silvia Trabucco}
\email[]{silvia.trabucco@gssi.it}
\affiliation{Gran Sasso Science Institute, Viale Francesco Crispi 7, 67100 L'Aquila, Italy} \affiliation{INFN Laboratori Nazionali del Gran Sasso, Via G. Acitelli 22, 67100 Assergi (AQ), Italy}

\begin{abstract}
\noindent We present a superfluid dark star model  consisting of relativistic  dark bosons with two-body self-interaction. The obtained masses, radii, and tidal deformability  depend in a simple way on the boson mass and  interaction strength. We report  first  results on binary mergers: the  distinctive amplitude and frequency of the emitted gravitational waves are well within reach of terrestrial interferometers. 
\end{abstract}

\maketitle

\noindent Boson stars~\cite{Kaup:1968zz, Ruffini:1969qy, Breit:1983nr} are hypotetical objects entirely made of bosons, see~\cite{Jetzer:1991jr, FranzESchunck_2003, Visinelli:2021uve, Shnir:2022lba} for  reviews. The idea originated from the work of Wheeler on geons~\cite{PhysRev.97.511}, appropriately adapted to Einstein-Klein-Gordon solitons.  
Indeed, in  the literature boson stars are mainly identified with localized soliton-like configurations  stabilized by gravity~\cite{PhysRev.172.1331, PhysRev.168.1445, PhysRev.187.1767, Friedberg:1986tq, PhysRevD.96.024002, Bettoni:2013zma}. They can be viewed as macroscopic quantum systems subject to their own gravitational attraction---the collapse is typically prevented by the Heisenberg uncertainty principle. In soliton-like stars, interactions significantly determine the stellar configuration but do not play an essential role in their gravitational stability.

In the present paper we discuss the equilibrium configuration of a fluid of self-interacting dark matter bosons subject to their own gravitational attraction. Besides the gravitational pull, bosons statistically attract each other, hence a short-range repulsion is needed to prevent the collapse.
Such approach parallels the observation that stable stars made of charged pions~\cite{Carignano:2016lxe,Brandt:2018bwq, Andersen:2018nzq, Mannarelli:2019hgn} may be realized if the pion electric charge is  balanced~\cite{Stashko:2023gnn}. 
In that case, the configuration is not a soliton, but can be viewed as a standard gas of self-gravitating matter. 

We focus on a simple system of dark bosons, completely decoupled from Standard Model.
We assume that the system features a global number symmetry that is spontaneously broken,  so that a superfluid is formed. The interparticle repulsion is modeled by a two-body self-interaction~\cite{Georgi:1985kw, peskin1995introduction}, allowing to determine analytically the equation of state (EoS).  
In the low-temperature limit, hydrostatically stable configurations, hereafter superfluid dark stars, have masses and radii linearly  scaling with the ratio between the square root of the self-coupling and the square of the boson mass. 
Since superfluid dark stars 
have arbitrarily low masses, 
they are compatible with the possible observation of low-mass compact objects. For instance, the component masses in the SSM200308 merger event~\cite{prunier2023analysis} are estimated  to be $M_1 = 0.62^{+0.46}_{-0.20}\,M_\odot$  and $M_2 = 0.27^{+0.12}_{-0.10}\, M_\odot$, with $M_\odot$  the solar mass.
Such low  masses  challenge the standard paradigm of hadronic compact stars~\cite{Shapiro-Teukolsky, Douchin:2001sv, Page:2006ud, Blaschke:2018mqw, Lattimer:2021emm, Burgio:2021vgk} originating from supernova explosion~\cite{Shapiro-Teukolsky, 1989ApJ...339..318C, Strobel:1999vn, Suwa:2018uni}. 

We address the tidal deformability and study numerically the  evolution of equal-mass binary systems, considering canonical  $1.4 M_\odot$, as well as lighter  $0.6 M_\odot$,  masses. We report the corresponding gravitational waveforms and amplitude spectral densities, as well as an estimate of the ejected matter.
We use natural units $\hbar = c= G_N=1$.

\SECTION{The model} We consider an ensemble of bosons at vanishing temperature and finite density. The system features a global $U(1)$ symmetry  associated to particle number conservation. 
We analyze the simplest model:  a massive complex scalar field, $\Phi$, having quartic self-interaction coupling $\lambda$~\cite{Friedberg:1986tq, Kleihaus:2005me, Kleihaus:2007vk}, with  $0 < \lambda \ll 1$. Such system was  considered in the context of solitonic boson stars for the first time in~\cite{Khlopov:1985, Colpi:1986ye}. The corresponding Lagrangian density can be written as
\be\label{eq:L_lambda}
{\cal L} =
D_\nu \Phi^* D^\nu \Phi - m^2_\bb |\Phi|^2 - \lambda |\Phi|^4\,,
\ee
where $m_\bb$ is the boson mass and the covariant derivative
\be D_\nu  = \de_\nu - i  \frac{\mu}\gamma u_\nu\,, \ee
appropriately includes the boson chemical potential $\mu$~\cite{Son:2002zn, Mannarelli:2008jq} and  the fluid four-velocity, $u^\nu = \gamma (1,{\bm u})$, with  $\gamma$ the Lorentz factor. In the non-relativistic limit, the above Lagrangian  is equivalent to the Gross-Pitaevskii one, which captures the main features of dilute  boson gases~\cite{Dalfovo:1999zz}. Superfluid vortices can be included quantizing the velocity field circulation; we comment on their possible relevance in the conclusions.

Upon expanding the covariant derivative, we obtain the potential minimum
\be
\label{eq:minimum}
 |\Phi|^2  = \begin{cases} \frac{1}{\gamma^2}\frac{\mu ^2 -m_\bb ^2\gamma^2}{2 \lambda} &  \text{for $|\mu| \geq m_\bb \gamma$} \\
 0 & \text{for $|\mu| < m_\bb \gamma$} \end{cases}\,,
 \ee
where $|\mu| = m_\bb \gamma$ corresponds to a second order quantum phase transition. 
At vanishing temperature, neglecting fluctuations, the quantum pressure in the superfluid phase 
\be \label{eq:pressure}
P(\mu)= \frac{(\mu ^2/\gamma^2 - m ^2_\bb)^2}{4\lambda} \, ,
\ee
is obtained by maximising the Lagrangian in Eq.~\eqref{eq:L_lambda}. 
 A non-vanishing velocity not only reduces the pressure, a relativistic effect analogous to the Bernoulli law, but it also decreases the order parameter.
 Since in the merging simulations the relativistic hydrodynamics is already taken into account, hereafter we consider the fluid at rest $u_\nu = \delta_{\nu 0}$.  The superfluid depletion due to relativistic effects  will be discussed in future work.

For a superfluid at rest, the quantum pressure vanishes at the transition point $|\mu| =m_\bb$, and it is independent  of the sign of $\mu$. From dimensional analysis $P \propto (\mu ^2 - m ^2_\bb)^2/\lambda$, where the $1/\lambda$ dependence reflects the depletion of the superfluid with increasing repulsion. These arguments can be easily generalized to include additional interactions, as in~\cite{Pitz:2023ejc},  which would generate pressure contributions  proportional to powers of $(\mu ^2 - m _\bb ^2)$. 

Using standard thermodynamic relations, we obtain the number density and the adiabatic speed of sound 
\be n = \mu \frac{\mu ^2 - m ^2_\bb}{\lambda} \quad \text{and} \quad c_s^2= \frac{\mu ^2 -m^2_\bb}{3 \mu ^2 - m^2_\bb}\,,\ee both vanishing at $|\mu|=m_\bb$.
 The speed of sound is a monotonic increasing function of the number density, approaching the conformal limit value for asymptotic densities. Since $c_s \leq 1/ \sqrt{3}$, the obtained EoS is quite soft compared to hadronic EoSs~\cite{Lattimer:2015nhk,  Page:2006ud, Blaschke:2018mqw, Lattimer:2021emm, Burgio:2021vgk}, which in many cases have speed of sound approaching the speed of light at large densities. Given $n$ and $P$, 
the  energy density is 
\begin{equation}
\epsilon =2 \frac{m^2_\bb}{\sqrt{\lambda}} \sqrt{P}+3P \,,
\label{eq:EoS}    
\end{equation}
which agrees with the effective EoS derived in \cite{Colpi:1986ye} for a fluid star and in \cite{PhysRevD.105.023001} for hybrid stars. The  {\it ansatz} in the approach of~\cite{Colpi:1986ye} is that $\Phi = \Phi_0(r) \exp (-i \omega t)$, hence we interpret  $\Phi_0(r)$ as the order parameter associated to the spontaneous symmetry breaking, while $\omega$ corresponds to the boson chemical potential.
With increasing $\lambda$, for fixed $\epsilon$, the pressure  grows, hence the   {\it stiffness} increases as the interparticle repulsion strengthens.

\SECTION{Hydrostatic equilibrium}
\label{sec:TOV}
Knowing the EoS,  we determine the hydrostatic equilibrium  configuration for  non-rotating mass distributions  by solving the spherical Tolman–Oppenheimer–Volkoff (TOV) equations~\cite{Tolman, Oppenheimer-Volkoff}.
To this end, the following rescaling is useful
\begin{align} 
\epsilon & = \frac{m ^4_\bb}{\lambda} \hat \epsilon  \, , \qquad 
P= \frac{m ^4_\bb}{\lambda} \hat P\,,
\label{eq:EpsandPscaling}\\
r &=  \frac{\sqrt{\lambda}}{m ^2_\bb} \hat r \,, \qquad   M= \frac{\sqrt{\lambda}}{m ^2_\bb} \hat M  \label{eq:RandMscaling}\,,
\end{align}
where $r$ is the radial distance from the origin and $M\equiv M(r)$ is the gravitational mass within $r$.
Both $\hat{\epsilon}$ and $\hat{P}$ are dimensionless in natural units. 
Upon substituting the above expressions in Eq.~\eqref{eq:EoS}, we have the dimensionless EoS
\be
\label{eq:eos}
\hat{ \epsilon} = 3 \hat{P} + 2 \sqrt{\hat{ P}}\,,
\ee
while the TOV equations are
\begin{equation}
\frac{d \hat{M}}{d \hat{r}}= 4 \pi \hat{r}^2 \hat{\epsilon }\,,\qquad
\frac{d \hat{P}}{d \hat{r}}= (\hat{\epsilon} + \hat{P})\frac{\hat{M} + 4 \pi \hat{r}^3 \hat{P}}{2\hat{M}  \hat{r} -\hat{r}^2}\,,
\end{equation}
which can be solved for a given central energy density, $\hat{\epsilon}_0 $.
Differently from solitonic-like stars~\cite{Jetzer:1991jr, FranzESchunck_2003, Visinelli:2021uve, Shnir:2022lba}, superfluid dark stars have a well defined radius, $R$, determined by the vanishing pressure condition~\cite{Shapiro-Teukolsky}. The gravitational mass is $M(R)$.
 \begin{figure}
\includegraphics[width=\columnwidth]{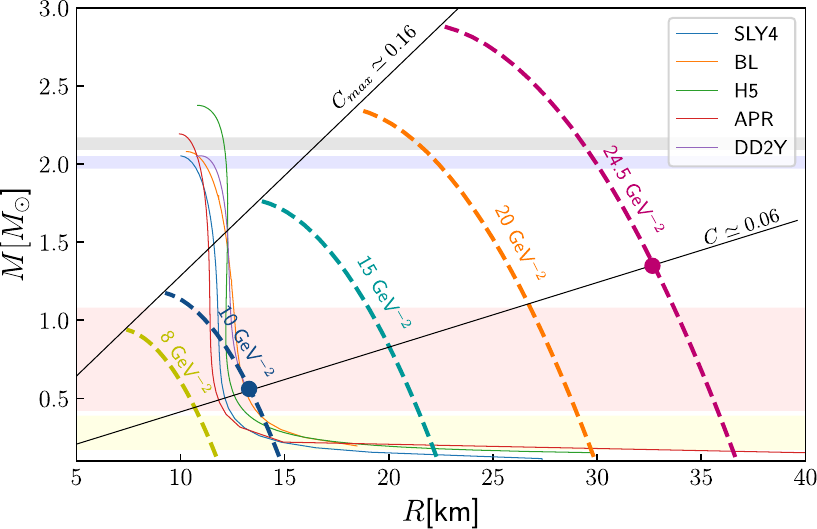}%
\vspace*{-1.5ex}
\caption{Masses and radii of superfluid dark stars for
different $x= \sqrt{\lambda}/m^2_\bb$ (dashed) compared with various hadronic EoSs (solid)~\cite{CHABANAT1998231, BL, Grams2022, PhysRevC.58.1804, PhysRevC.81.015803}.
The top gray (blue) band corresponds to the observational limits on pulsar masses in~\cite{Romani_2021}(\cite{Antoniadis_2013}). The bottom pink and yellow bands show the component masses in the SSM200308 event~\cite{prunier2023analysis}. 
The dots correspond to the configurations used in the merger simulations.
The two thin straight lines indicate the maximum compactness (upper line) and the one used in the merger simulations (lower line).}
\label{fig:MR}
\end{figure}

In Fig.~\ref{fig:MR} we show the mass-radius diagrams of the superfluid dark stars for different values of 
\be\label{eq:x}
x=\frac{\sqrt{\lambda}}{m^2_\bb} \,,
\ee
which is the parameter determining the scaling of masses and radii in Eqs.~\eqref{eq:RandMscaling}. We compare the results with those obtained using few hadronic EoSs.
For given $x$, lowering the central density, the mass decreases while the radius increases.  In the low density limit, the EoS~\eqref{eq:EoS} is  approximated by the polytrope $P = \frac{x^2}{4} \epsilon^2$
so that the radius is given by $R = x \sqrt{\frac{\pi}{8}}$, which is confirmed in Fig.~\ref{fig:MR} for vanishing $M$. We also find that at low densities $M$ scales linearly with the central density.

The maximal mass of superfluid dark stars scales linearly with $x$,  reaching the neutron star (NS) observational limit of about $2.1 M_\odot$, for $x \simeq 20 $ GeV$^{-2}$. Compared to hadronic EoSs, the corresponding stellar radius is here about a factor $2$ larger.   Since both  $R$ and $M$  scale linearly  with  $x$, extremely massive superfluid dark stars  have  radii much larger than standard NSs.
On the other extreme, if $x$ is very small,  light clumps of superfluid dark matter with very small radii may be  realized.  This would happen for small values of $\lambda$ and/or for heavy dark bosons, namely large $m_\bb$.

\begin{figure}
\includegraphics[width=0.95\columnwidth]{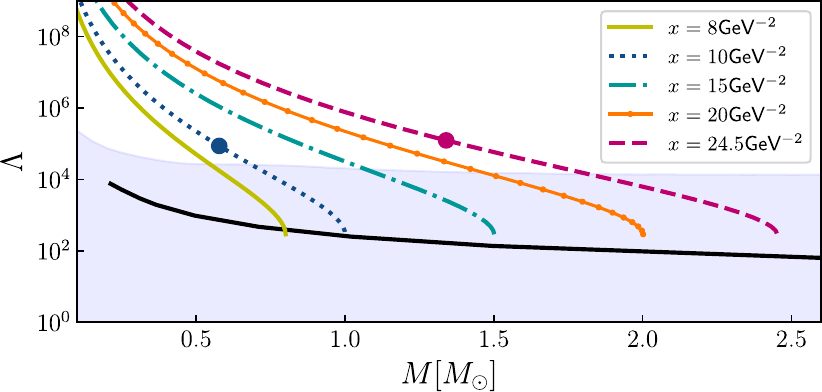}%
 \vspace*{-1.5ex}
\caption{Dimensionless tidal deformability $\Lambda$, see Eq.~\eqref{eq:lambda}, as a function of the stellar gravitational mass $M$.  For any $x$ in Eq.~\eqref{eq:x}, the curves are obtained changing the dimensionless central energy density. The lowest value $\Lambda_\text{min} \simeq 290$ is reached for the maximal mass, while  $\Lambda$ diverges for low masses. The dots correspond to the configurations used in the merger simulations.  The shaded area delimits the $3\sigma$ upper bound at fixed signal to noise ratio SNR$=12$ in the O4 observational run in LVK~\cite{Pani, LIGO-sens, crescimbeni2024primordial}. The black line corresponds to the $1\sigma$ tidal deformability at $100\,\unit{Mpc}$ for the $\Delta-10$ km HFLF-Cryo configuration of ET~\cite{Branchesi:2023mws}.  }
\label{fig:lambda}
\end{figure}

The thin straight lines in  Fig.~\ref{fig:MR}, indicate two different values of  the  stellar compactness, $\mathcal{C}= M/R$.  Given the scaling in Eq.~\eqref{eq:RandMscaling}, the compactness  is independent of $x$, therefore these lines join configurations with the same dimensionless central energy density but different $x$. Since $M \propto \hat M \, x$ and  $\epsilon \propto \hat \epsilon\, x^{-2}$, stars having the same compactness but larger masses (and radii) have smaller central densities.
The   maximum   compactness is ${\cal C }_{\rm max}\simeq 0.16$, which is smaller than the  values obtained with hadronic matter---the dark boson EoS is relatively soft. Thanks to  this upper limit, future evidence of higher compactness may  rule out superfluid dark stars obtained with~\eqref{eq:L_lambda}.  Higher compactness can be obtained with different contact interactions~\cite{Pitz:2023ejc}.
Such a measurement should result from gravitational observations since dark bosons  are not expected to emit light. Hence, bounds from NICER~\cite{Riley:2019yda} as well as from GW170817~\cite{TheLIGOScientific:2017qsa}, which was followed by  a  kilonova~\cite{LIGOScientific:2017ync}, cannot constrain our model.

Another possibility is to assess the existence of superfluid dark stars by tidal deformation measurements.
We calculate  the so-called tidal Love number $k_2$  as in~\cite{Hinderer:2007mb, Flanagan:2007ix, Damour:2009vw, Postnikov:2010yn}: it ranges from about $0.045$ to $0.25$ for decreasing compactness. In Fig.~\ref{fig:lambda}, we show  the mass dependence of the dimensionless tidal deformability parameter 
\begin{equation}\label{eq:lambda}
\Lambda = \frac{2}{3 \mathcal{C}^{5}}\,k_2\,,
\end{equation}
together with the expected reach of the O4 run at Ligo-Virgo-Kagra (LVK)~\cite{Pani, LIGO-sens} and the estimated sensitivity 
of the Einstein Telescope (ET)~\cite{Branchesi:2023mws}.  We observe that for each value of $x$ there is a range of masses for which a measurement should be possible. In particular,  ET should be able to probe the entire range of tidal deformabilities of superfluid dark stars.
For vanishing masses, all $\Lambda$ curves in Fig.~\ref{fig:lambda} diverge. Towards maximal $M$, the lowest value $\Lambda_\text{min}\simeq 290$ is obtained, regardless of $x$, consistent with the fact that the external part of the star can be described by a  
$\Gamma \simeq 2$ polytrope~\cite{Postnikov:2010yn}.
The two dots correspond to the stars used in the merger simulations. Since these two configurations have similar dimensionless central densities $\hat{\epsilon}_0 \sim 10^{-4}$, they also have similar tidal deformabilities~$\sim 10^5$. We remark that as Einstein's equations for a superfluid dark star can be written in dimensionless units with the appropriate rescaling of Eqs.~\eqref{eq:EpsandPscaling} and~\eqref{eq:RandMscaling}, we expect that the value of any dimensionless observable, e.g. the tidal deformability, at fixed central density, $\hat \epsilon_0$, should be independent of $x$. 

\medskip

\SECTION{Binary Mergers}
Considering dark bosons  decoupled from Standard Model particles, the only way to probe superfluid dark stars is by their gravitational effects~\cite{Bertone:2019irm}, see however the discussion in~\cite{Rosa:2022tfv} for possible light emission mechanisms. In the context of soliton-like models, merging of boson stars was considered in~\cite{Palenzuela:2007dm, Bezares:2018qwa, PhysRevD.96.104058}. Here we study the merging of two equal-mass inspiraling superfluid dark stars, with compactness  $\mathcal{C} \simeq 0.06$,  corresponding to the two dots in  Figs.~\ref{fig:MR} and~\ref{fig:lambda}. The first simulation is done with  stars having constituent  mass\,\footnote{This is the stellar mass computed integrating the energy density
over the proper volume.} $M_b \simeq 1.4\,M_\odot$ and radius $R\simeq 32.5\,\text{km}$ at an initial distance of $90\,\text{km}$, while the second has stars with $M_b \simeq 0.6\,M_\odot$ and $R\simeq13.1\,\text{km}$ at an initial distance of $50\,\text{km}$. 

 The numerical setup employed is analogous to that of~\cite{DePietri:2019mti}, in which initial data for irrotational binaries are generated with \textsc{LORENE}~\cite{loreneWS, Gourgoulhon_2001} and evolved dynamically using the \textsc{Einstein Toolkit}~\cite{L_ffler_2012}. The spatial resolution of the grid is chosen to reasonably represent the interior of the stars while ensuring that the computational cost remains manageable: production runs have been performed with $\delta=553\,\text{m}$.

\begin{figure}
  \centering
 \includegraphics[width=0.98\columnwidth]{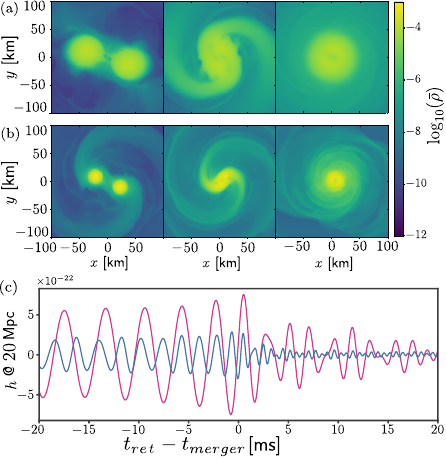}%
 \vspace*{-1.5ex}
  \caption{Results of the merger simulations.  (a-b) Three snapshots of the matter density distribution  $\bar{\rho} = \rho\,G^{3}M_\odot^{2}c^{-6}$, taken during inspiral, merger and postmerger for  $M_b= 1.4\,M_\odot$ (a), and  $M_b = 0.6\,M_\odot$ (b).  
  (c) The extracted GW signals in magenta and blue corresponding to (a) and (b), respectively.} 
  \label{fig:sims}
\end{figure}

Fig.~\ref{fig:sims} shows a summary of the two simulations. None of the systems collapses to a black hole in the simulated time, as expected from the low value of the compactness. The top panels contain snapshots of the density distribution at different times; the bottom one shows the extracted $h_{22}$ component of the gravitational wave (GW)  in magenta  and blue for the higher and lower mass respectively-- see~\cite{Hinder_2013} for detailed extraction procedure.

\begin{figure}
   \includegraphics[width=\columnwidth]{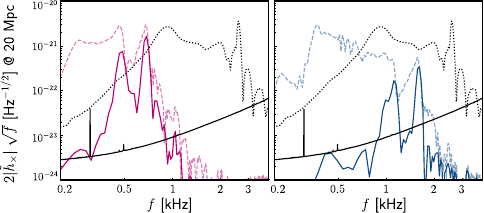}%
 \vspace*{-1.5ex}
   \caption{Amplitude spectral densities $2| \tilde{h}_\times |\sqrt{f}$ (see e.g.~\cite{Rezzolla:2016nxn}) at $20\unit{Mpc}$ for binary merging of equal mass stars, for $M=1.4(0.6)\,M_\odot$ respectively in the left (right) panel and magenta (blue) lines. The dotted black lines show the spectrum for DD2 EoS~\cite{Bauswein2016}. The solid black lines are the sensitivity curve of Advanced LIGO~\cite{LIGO-sens}.}
   \label{fig:spectra}
\end{figure}

For $M_b =1.4\,M_\odot$, the magenta line in Fig.~\ref{fig:sims}(c) shows the typical chirp, followed by the merger and the formation of a bar-deformed remnant. Its imprint on the GW can be seen in the spectral density Fig.~\ref{fig:spectra}(left), where two dominant frequencies appear: one at $f \simeq 475\,\text{Hz}$ and the other at $f \simeq 670\,\text{Hz}$. They turn on in the postmerger phase, as shown by the spectrum of the waveform computed after $t_{\text{merger}}$. We confirm that the peaks originate from the $m=1$ and $m=2$ matter oscillation modes, see e.g.~\cite{Stergioulas:2011}.

For $M_b=0.6\,M_\odot$, a more peculiar behaviour emerges, see the blue line in Fig.~\ref{fig:sims}(c). Somewhat evident are the oscillations due to the residual eccentricity in the initial data. They are expected to be  pronounced for soft EoS models and small stellar masses. In order to test the reliability of the method, we estimated the residual eccentricity in the Newtonian approximation (following the distance of one star from the origin, see for instance~\cite{PhysRevD.90.064006, Dietrich_2015}). The result is about $5 \times 10^{-2}$,  compatible with residuals from initial data. 
The right panel in Fig.~\ref{fig:spectra} shows the corresponding  spectrum. It is  qualitatively similar to the left one; the two peaks are only shifted to higher frequencies, $f\simeq 1.1\,\text{kHz}$ and $f\simeq 1.54\,\text{kHz}$, and appear less intense due to the shorter GW signal. It is encouraging to see that in both simulations the  peak frequencies  are well within the observational window of LVK~\cite{LVK}, and for high mass objects are at lower frequencies with respect to NS mergers~\cite{Bauswein2016, PhysRevD.86.063001, Pani:private}.

The obtained results confirm that for fixed compactness 
observables scale  with $x$. The ratio of the initial masses (or equivalently, initial radii) is equal to the ratio of the used values of $x$ (in this case, $2.45$).
The spectral pattern is peaked at frequencies that increase with decreasing $x$: the peaks in the two plots of Fig.~\ref{fig:spectra} have a  ratio  $\simeq 2.3$. A  similar value comes from the minimization of the $L_2$ norm between the two strains of Fig.~\ref{fig:sims}(c), that is given by
\be
\argmin_{d} \| h_{1.4} - d\, h_{0.6}\|_2 \simeq 2.3\,,
\ee
where $d$ is  a scaling factor. A relative error of about $6\%$  may be  in part due to differences in the initial data (the exact values of the two compactness differ by $\sim4\%$) and in part to numerical errors accumulated during the simulations, the extraction procedure and the Fourier transform.

The formation of a disk of ejected matter surrounding the remnant is visible in both Fig.~\ref{fig:sims}(a)-(b). Its constituent mass can be computed as the three-dimensional integral of the conserved rest-mass density $D=\sqrt{g}\bar{\rho} \gamma$ \cite{1997ApJ...476..221B} from the minimum value to some upper limit $D_\text{cut}$. Here $g$ is the determinant of the 3-metric $g_{ij}$, $\gamma$ the Lorentz factor and $\bar{\rho}$ the density in geometrized units. However, there is no unambiguous way to choose the threshold to separate the remnant from the disk.

\begin{figure}
   \includegraphics[width=\columnwidth]{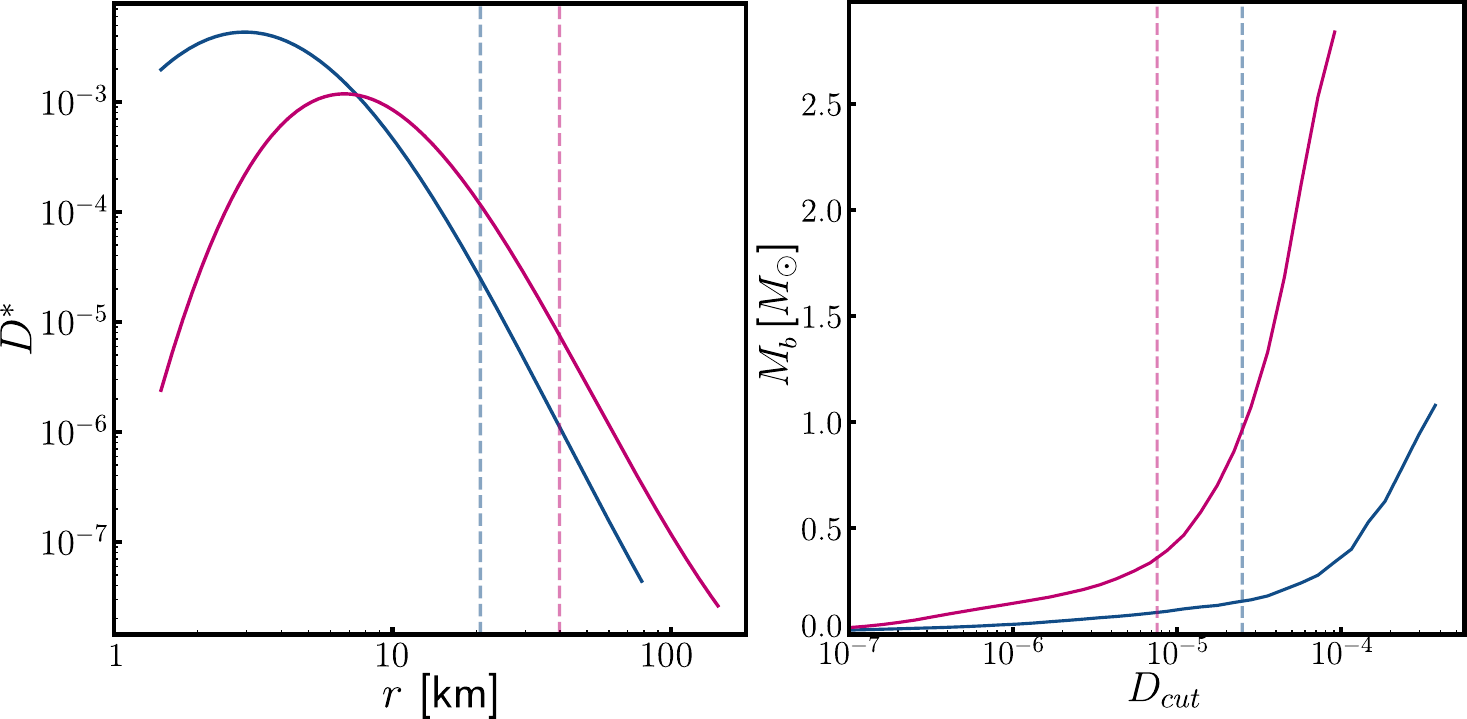}%
 \vspace*{-1.5ex}
   \caption{Left: azimuthal average of  $\bar{\rho}$ as function of the radial distance $r$. The dashed line represents the radius at which its logarithmic derivative is $-4.53$ for $M_b=1.4\,M_{\odot}$ in magenta and $M_b=0.6\,M_{\odot}$ in blue. Right: total constituent mass computed integrating densities up to $D_\text{cut}$. }
   \label{fig:disk}
\end{figure}

To this end,  we look at the density profile of the fluid on the XY plane at a time in which the mass density reaches a steady state and the $m=2$ deformation is subdominant with respect to the overall axial symmetry. Fig.~\ref{fig:disk}(left) shows, for both simulations, its average over the entire azimuthal angle
\be
\label{eq:v*}
D^*(r) = \frac{1}{2\pi}\int_{-\pi}^{\pi}\!\!\!D\,\rm{d}\phi\,,
\ee
in magenta for $M_b = 1.4 \,M_\odot$ and in blue for $M_b = 0.6 \,M_\odot$. Looking at this plot together with the third panel of Fig.~\ref{fig:sims}(a) and (b), it is possible to distinguish the main features of the remnant: in the centre there is a region of lower density around which two clumps of matter rotate, slowly dissipating energy (in our simulations only via GW emission). We set the density threshold looking for the distance at which the logarithmic derivative of the density is
\be
\frac{\rm{d}\log(D^*)}{\rm{d}\log(r)} = -4.53\,,
\ee
where the numerical value guarantees that disk masses in the two simulations satisfy the $x$--scaling. In Fig.~\ref{fig:disk}(right) we show how the computed disk mass varies when the upper limit in the density integration is changed, and where our estimates lie. The values found are marked as dashed vertical lines. As can be seen from the right panel of Fig.~\ref{fig:disk}, they approximately capture the change in slope of the mass profiles. The corresponding disk masses   are $M_{\text{disk}} \simeq 0.36 \,M_\odot$ and $M_{\text{disk}} \simeq 0.16 \,M_\odot$, respectively. They are about $13\%$ of the initial total mass.

\SECTION{Conclusions and outlook}
We presented a  star model composed of  self-gravitating superfluid dark matter, showing that  its stellar structure  depends on just two parameters: the central density and the $x$ parameter given by  the ratio of the  square root of the coupling to the square boson mass. The observational $\sim 2.1 M_\odot$ constraint on the compact star mass is compatible with our findings only if their radius  is larger than $\sim 20$ km. Large and massive  superfluid dark stars may be realized for large value of the $x$ parameter, see Eq.~\eqref{eq:x}.

Given that superfluid dark stars are not expected to emit  light and neutrinos as standard NSs, the only  viable way to assess their existence and constraint their masses, radii and tidal deformations is by gravitational  observations. In this respect, we have reported  a  study of superfluid dark-stars merger, showing that for relatively low masses the final object is gravitationally stable. The GW signal is quite different from that of  hadronic star mergers and possibly detectable by LVK. 
 
This analysis  can be  extended and improved in various different ways.
We could simulate  merging events of two very massive superfluid dark stars,  having  masses  of $20 M_\odot$, or even larger,  and then compare  the GW signal with that of a black hole merger.  For such large masses,  the uncertainty in the determination of the tidal deformability in ET~\cite{Branchesi:2023mws} should be low enough to allow to distinguish superfluid dark stars from black holes.
Hybrid compact stars in which standard hadronic matter  coexists with superfluid  dark matter  may be realized, as well.  Such hybrid models have been already  discussed in several cases~ \cite{Goldman:1989nd, Bertone:2007ae, Ciarcelluti:2010ji, Li:2012qf, Ellis:2018bkr, Nelson:2018xtr, PhysRevD.102.063028, PhysRevD.105.023001, Das:2020vng, Berezhiani:2020zck, Giangrandi:2022wht, Kumar:2022amh, PhysRevD.109.043029, Thakur:2023aqm}, in   particular dark matter in the Bose-Einstein condensed BEC phase has been considered in \cite{Li:2012qf, Giangrandi:2022wht, PhysRevD.105.023001, PhysRevD.109.043029}, showing a softening of the EoS as a consequence of the  dark matter presence.

The results of the merging simulation show that observables scale with $x$. However, it is worth remarking that we obtained such results for fixed $m_\bb$. Varying the value of the boson mass while keeping $x$ constant should   produce a number of effects. The particle velocities are expected to change
resulting 
in the condensate depletion. Rotating superfluids host quantized vortices with core size of the order 
$\xi \simeq ( \sqrt{2} m_\bb c_s)^{-1}$~\cite{Dalfovo:1999zz}, while  the velocity circulation 
is quantized in units of $1/m_\bb $. Changing $m_\bb$, the whole vortex structure changes, hence, rotation properties of superfluid dark stars are not scalable with $x$ 
but explicitly depend on the boson mass. For sufficiently low $m_\bb$, rotating boson stars are expected to host a small number of very large vortices. In this case, the  phenomenology can be forecast from cold atom experiments; for instance 
vortex core precession could take place as observed in ultracold dilute gases~\cite{Anderson_2000}. 

Since in this model it is possible to have  light  dark matter lumps, heavy superfluid dark stars may result from sequential merging of small droplets. 
Clearly, temperature effects should be included and may lead to new phenomena~\cite{Dolan:1973qd}. 
However, for virialized  dark stars we expect that $T \propto m_\text{B} \cal{C}$, while the BEC  
 critical temperature is $T_c \propto n^{2/3}/m_\text{B}$ (corrections due to interactions are negligible for sufficiently small $\lambda$ \cite{Stoof:1992zz,1997PhRvL..79.3549G}), thus (at fixed $x$) the condition $T \ll T_C$ is satisfied for sufficiently small values the  boson mass. Upon substituting the numerical values of the parameters we find an upper bound on the boson mass of hundreds of MeV.
During the binary merging, due to the high energies involved, part of matter could evaporate. In principle, if the two colliding objects are sufficiently light, they may evaporate even before the merging is concluded. Hence, arbitrary small dark lumps of condensed  bosons are not expected to form unless they are ultracold  and survive encounters.  In this respect, it would be interesting to study whether viscous effects~\cite{PhysRevD.53.5799} may influence the thermal and hydrodynamic evolution of merging dark droplets. One should  analyze their collisions  to infer whether they merge or dissolve.  In the latter case they may contribute to the cold  dark matter component.  Indeed, self-interacting dark matter happens to be a good candidate to alleviate some of dark-matter problems~\cite{Spergel:1999mh, Tulin:2017ara}, see as well the discussion in~\cite{Ianni:2021ynp, Nesti:2023tid}.  
A guidance in the study of dark droplet interactions can derive from studies of colliding ultracold atom droplets~\cite{PhysRevLett.122.090401}, which result in either merging or separation depending on their incoming relative velocity.

If the merging droplets of dark bosons  are sufficiently compact and cold, they may create tiny black holes. This would be of great astrophysical interest due to the quick evaporation  by Hawking emission,  resulting  in a mechanism to turn dark matter into Standard Model  particles. The present model may be extended to include appropriate self-interactions that drive the system in a supersolid phase~\cite{norcia2021tds,Bland2022tds,norcia2022cao,klaus2022oov}, and rotating supersolid dark stars may be quantum simulated as in~\cite{Poli:2023vyp}. 

\medskip

\begin{acknowledgments}
\emph{Acknowledgments.} We thank Paolo Pani, Marica Branchesi and Roberto De Pietri for useful discussions.
\end{acknowledgments}


\begin{thebibliography}{109}%
\makeatletter
\providecommand \@ifxundefined [1]{%
 \@ifx{#1\undefined}
}%
\providecommand \@ifnum [1]{%
 \ifnum #1\expandafter \@firstoftwo
 \else \expandafter \@secondoftwo
 \fi
}%
\providecommand \@ifx [1]{%
 \ifx #1\expandafter \@firstoftwo
 \else \expandafter \@secondoftwo
 \fi
}%
\providecommand \natexlab [1]{#1}%
\providecommand \enquote  [1]{``#1''}%
\providecommand \bibnamefont  [1]{#1}%
\providecommand \bibfnamefont [1]{#1}%
\providecommand \citenamefont [1]{#1}%
\providecommand \href@noop [0]{\@secondoftwo}%
\providecommand \href [0]{\begingroup \@sanitize@url \@href}%
\providecommand \@href[1]{\@@startlink{#1}\@@href}%
\providecommand \@@href[1]{\endgroup#1\@@endlink}%
\providecommand \@sanitize@url [0]{\catcode `\\12\catcode `\$12\catcode `\&12\catcode `\#12\catcode `\^12\catcode `\_12\catcode `\%12\relax}%
\providecommand \@@startlink[1]{}%
\providecommand \@@endlink[0]{}%
\providecommand \url  [0]{\begingroup\@sanitize@url \@url }%
\providecommand \@url [1]{\endgroup\@href {#1}{\urlprefix }}%
\providecommand \urlprefix  [0]{URL }%
\providecommand \Eprint [0]{\href }%
\providecommand \doibase [0]{http://dx.doi.org/}%
\providecommand \selectlanguage [0]{\@gobble}%
\providecommand \bibinfo  [0]{\@secondoftwo}%
\providecommand \bibfield  [0]{\@secondoftwo}%
\providecommand \translation [1]{[#1]}%
\providecommand \BibitemOpen [0]{}%
\providecommand \bibitemStop [0]{}%
\providecommand \bibitemNoStop [0]{.\EOS\space}%
\providecommand \EOS [0]{\spacefactor3000\relax}%
\providecommand \BibitemShut  [1]{\csname bibitem#1\endcsname}%
\let\auto@bib@innerbib\@empty
\bibitem [{\citenamefont {Kaup}(1968{\natexlab{a}})}]{Kaup:1968zz}%
  \BibitemOpen
  \bibfield  {author} {\bibinfo {author} {\bibfnamefont {D.~J.}\ \bibnamefont {Kaup}},\ }\href {\doibase 10.1103/PhysRev.172.1331} {\bibfield  {journal} {\bibinfo  {journal} {Phys. Rev.}\ }\textbf {\bibinfo {volume} {172}},\ \bibinfo {pages} {1331} (\bibinfo {year} {1968}{\natexlab{a}})}\BibitemShut {NoStop}%
\bibitem [{\citenamefont {Ruffini}\ and\ \citenamefont {Bonazzola}(1969{\natexlab{a}})}]{Ruffini:1969qy}%
  \BibitemOpen
  \bibfield  {author} {\bibinfo {author} {\bibfnamefont {R.}~\bibnamefont {Ruffini}}\ and\ \bibinfo {author} {\bibfnamefont {S.}~\bibnamefont {Bonazzola}},\ }\href {\doibase 10.1103/PhysRev.187.1767} {\bibfield  {journal} {\bibinfo  {journal} {Phys. Rev.}\ }\textbf {\bibinfo {volume} {187}},\ \bibinfo {pages} {1767} (\bibinfo {year} {1969}{\natexlab{a}})}\BibitemShut {NoStop}%
\bibitem [{\citenamefont {Breit}\ \emph {et~al.}(1984)\citenamefont {Breit}, \citenamefont {Gupta},\ and\ \citenamefont {Zaks}}]{Breit:1983nr}%
  \BibitemOpen
  \bibfield  {author} {\bibinfo {author} {\bibfnamefont {J.~D.}\ \bibnamefont {Breit}}, \bibinfo {author} {\bibfnamefont {S.}~\bibnamefont {Gupta}}, \ and\ \bibinfo {author} {\bibfnamefont {A.}~\bibnamefont {Zaks}},\ }\href {\doibase 10.1016/0370-2693(84)90764-0} {\bibfield  {journal} {\bibinfo  {journal} {Phys. Lett.}\ }\textbf {\bibinfo {volume} {B140}},\ \bibinfo {pages} {329} (\bibinfo {year} {1984})}\BibitemShut {NoStop}%
\bibitem [{\citenamefont {Jetzer}(1992)}]{Jetzer:1991jr}%
  \BibitemOpen
  \bibfield  {author} {\bibinfo {author} {\bibfnamefont {P.}~\bibnamefont {Jetzer}},\ }\href {\doibase 10.1016/0370-1573(92)90123-H} {\bibfield  {journal} {\bibinfo  {journal} {Phys. Rept.}\ }\textbf {\bibinfo {volume} {220}},\ \bibinfo {pages} {163} (\bibinfo {year} {1992})}\BibitemShut {NoStop}%
\bibitem [{\citenamefont {Schunck}\ and\ \citenamefont {Mielke}(2003)}]{FranzESchunck_2003}%
  \BibitemOpen
  \bibfield  {author} {\bibinfo {author} {\bibfnamefont {F.~E.}\ \bibnamefont {Schunck}}\ and\ \bibinfo {author} {\bibfnamefont {E.~W.}\ \bibnamefont {Mielke}},\ }\href {\doibase 10.1088/0264-9381/20/20/201} {\bibfield  {journal} {\bibinfo  {journal} {Classical and Quantum Gravity}\ }\textbf {\bibinfo {volume} {20}},\ \bibinfo {pages} {R301} (\bibinfo {year} {2003})}\BibitemShut {NoStop}%
\bibitem [{\citenamefont {Visinelli}(2021)}]{Visinelli:2021uve}%
  \BibitemOpen
  \bibfield  {author} {\bibinfo {author} {\bibfnamefont {L.}~\bibnamefont {Visinelli}},\ }\href {\doibase 10.1142/S0218271821300068} {\bibfield  {journal} {\bibinfo  {journal} {Int. J. Mod. Phys. D}\ }\textbf {\bibinfo {volume} {30}},\ \bibinfo {pages} {2130006} (\bibinfo {year} {2021})},\ \Eprint {http://arxiv.org/abs/2109.05481} {arXiv:2109.05481 [gr-qc]} \BibitemShut {NoStop}%
\bibitem [{\citenamefont {Shnir}(2023)}]{Shnir:2022lba}%
  \BibitemOpen
  \bibfield  {author} {\bibinfo {author} {\bibfnamefont {Y.}~\bibnamefont {Shnir}},\ }\href {\doibase 10.1007/978-3-031-31520-6_10} {\bibfield  {journal} {\bibinfo  {journal} {Lect. Notes Phys.}\ }\textbf {\bibinfo {volume} {1017}},\ \bibinfo {pages} {347} (\bibinfo {year} {2023})},\ \Eprint {http://arxiv.org/abs/2204.06374} {arXiv:2204.06374 [gr-qc]} \BibitemShut {NoStop}%
\bibitem [{\citenamefont {Wheeler}(1955)}]{PhysRev.97.511}%
  \BibitemOpen
  \bibfield  {author} {\bibinfo {author} {\bibfnamefont {J.~A.}\ \bibnamefont {Wheeler}},\ }\href {\doibase 10.1103/PhysRev.97.511} {\bibfield  {journal} {\bibinfo  {journal} {Phys. Rev.}\ }\textbf {\bibinfo {volume} {97}},\ \bibinfo {pages} {511} (\bibinfo {year} {1955})}\BibitemShut {NoStop}%
\bibitem [{\citenamefont {Kaup}(1968{\natexlab{b}})}]{PhysRev.172.1331}%
  \BibitemOpen
  \bibfield  {author} {\bibinfo {author} {\bibfnamefont {D.~J.}\ \bibnamefont {Kaup}},\ }\href {\doibase 10.1103/PhysRev.172.1331} {\bibfield  {journal} {\bibinfo  {journal} {Phys. Rev.}\ }\textbf {\bibinfo {volume} {172}},\ \bibinfo {pages} {1331} (\bibinfo {year} {1968}{\natexlab{b}})}\BibitemShut {NoStop}%
\bibitem [{\citenamefont {Feinblum}\ and\ \citenamefont {McKinley}(1968)}]{PhysRev.168.1445}%
  \BibitemOpen
  \bibfield  {author} {\bibinfo {author} {\bibfnamefont {D.~A.}\ \bibnamefont {Feinblum}}\ and\ \bibinfo {author} {\bibfnamefont {W.~A.}\ \bibnamefont {McKinley}},\ }\href {\doibase 10.1103/PhysRev.168.1445} {\bibfield  {journal} {\bibinfo  {journal} {Phys. Rev.}\ }\textbf {\bibinfo {volume} {168}},\ \bibinfo {pages} {1445} (\bibinfo {year} {1968})}\BibitemShut {NoStop}%
\bibitem [{\citenamefont {Ruffini}\ and\ \citenamefont {Bonazzola}(1969{\natexlab{b}})}]{PhysRev.187.1767}%
  \BibitemOpen
  \bibfield  {author} {\bibinfo {author} {\bibfnamefont {R.}~\bibnamefont {Ruffini}}\ and\ \bibinfo {author} {\bibfnamefont {S.}~\bibnamefont {Bonazzola}},\ }\href {\doibase 10.1103/PhysRev.187.1767} {\bibfield  {journal} {\bibinfo  {journal} {Phys. Rev.}\ }\textbf {\bibinfo {volume} {187}},\ \bibinfo {pages} {1767} (\bibinfo {year} {1969}{\natexlab{b}})}\BibitemShut {NoStop}%
\bibitem [{\citenamefont {Friedberg}\ \emph {et~al.}(1987)\citenamefont {Friedberg}, \citenamefont {Lee},\ and\ \citenamefont {Pang}}]{Friedberg:1986tq}%
  \BibitemOpen
  \bibfield  {author} {\bibinfo {author} {\bibfnamefont {R.}~\bibnamefont {Friedberg}}, \bibinfo {author} {\bibfnamefont {T.~D.}\ \bibnamefont {Lee}}, \ and\ \bibinfo {author} {\bibfnamefont {Y.}~\bibnamefont {Pang}},\ }\href {\doibase 10.1103/PhysRevD.35.3658} {\bibfield  {journal} {\bibinfo  {journal} {Phys. Rev. D}\ }\textbf {\bibinfo {volume} {35}},\ \bibinfo {pages} {3658} (\bibinfo {year} {1987})}\BibitemShut {NoStop}%
\bibitem [{\citenamefont {Sennett}\ \emph {et~al.}(2017)\citenamefont {Sennett} \emph {et~al.}}]{PhysRevD.96.024002}%
  \BibitemOpen
  \bibfield  {author} {\bibinfo {author} {\bibfnamefont {N.}~\bibnamefont {Sennett}} \emph {et~al.},\ }\href {\doibase 10.1103/PhysRevD.96.024002} {\bibfield  {journal} {\bibinfo  {journal} {Phys. Rev. D}\ }\textbf {\bibinfo {volume} {96}},\ \bibinfo {pages} {024002} (\bibinfo {year} {2017})}\BibitemShut {NoStop}%
\bibitem [{\citenamefont {Bettoni}\ \emph {et~al.}(2014)\citenamefont {Bettoni}, \citenamefont {Colombo},\ and\ \citenamefont {Liberati}}]{Bettoni:2013zma}%
  \BibitemOpen
  \bibfield  {author} {\bibinfo {author} {\bibfnamefont {D.}~\bibnamefont {Bettoni}}, \bibinfo {author} {\bibfnamefont {M.}~\bibnamefont {Colombo}}, \ and\ \bibinfo {author} {\bibfnamefont {S.}~\bibnamefont {Liberati}},\ }\href {\doibase 10.1088/1475-7516/2014/02/004} {\bibfield  {journal} {\bibinfo  {journal} {JCAP}\ }\textbf {\bibinfo {volume} {02}},\ \bibinfo {pages} {004} (\bibinfo {year} {2014})},\ \Eprint {http://arxiv.org/abs/1310.3753} {arXiv:1310.3753 [astro-ph.CO]} \BibitemShut {NoStop}%
\bibitem [{\citenamefont {Carignano}\ \emph {et~al.}(2017)\citenamefont {Carignano}, \citenamefont {Lepori}, \citenamefont {Mammarella}, \citenamefont {Mannarelli},\ and\ \citenamefont {Pagliaroli}}]{Carignano:2016lxe}%
  \BibitemOpen
  \bibfield  {author} {\bibinfo {author} {\bibfnamefont {S.}~\bibnamefont {Carignano}}, \bibinfo {author} {\bibfnamefont {L.}~\bibnamefont {Lepori}}, \bibinfo {author} {\bibfnamefont {A.}~\bibnamefont {Mammarella}}, \bibinfo {author} {\bibfnamefont {M.}~\bibnamefont {Mannarelli}}, \ and\ \bibinfo {author} {\bibfnamefont {G.}~\bibnamefont {Pagliaroli}},\ }\href {\doibase 10.1140/epja/i2017-12221-x} {\bibfield  {journal} {\bibinfo  {journal} {Eur. Phys. J. A}\ }\textbf {\bibinfo {volume} {53}},\ \bibinfo {pages} {35} (\bibinfo {year} {2017})},\ \Eprint {http://arxiv.org/abs/1610.06097} {arXiv:1610.06097 [hep-ph]} \BibitemShut {NoStop}%
\bibitem [{\citenamefont {Brandt}\ \emph {et~al.}(2018)\citenamefont {Brandt}, \citenamefont {Endr\ifmmode~\mbox{\H{o}}\else \H{o}\fi{}di}, \citenamefont {Fraga}, \citenamefont {Hippert}, \citenamefont {Schaffner-Bielich},\ and\ \citenamefont {Schmalzbauer}}]{Brandt:2018bwq}%
  \BibitemOpen
  \bibfield  {author} {\bibinfo {author} {\bibfnamefont {B.~B.}\ \bibnamefont {Brandt}}, \bibinfo {author} {\bibfnamefont {G.}~\bibnamefont {Endr\ifmmode~\mbox{\H{o}}\else \H{o}\fi{}di}}, \bibinfo {author} {\bibfnamefont {E.~S.}\ \bibnamefont {Fraga}}, \bibinfo {author} {\bibfnamefont {M.}~\bibnamefont {Hippert}}, \bibinfo {author} {\bibfnamefont {J.}~\bibnamefont {Schaffner-Bielich}}, \ and\ \bibinfo {author} {\bibfnamefont {S.}~\bibnamefont {Schmalzbauer}},\ }\href {\doibase 10.1103/PhysRevD.98.094510} {\bibfield  {journal} {\bibinfo  {journal} {Phys. Rev. {\bf D}}\ }\textbf {\bibinfo {volume} {98}},\ \bibinfo {pages} {094510} (\bibinfo {year} {2018})}\BibitemShut {NoStop}%
\bibitem [{\citenamefont {Andersen}\ and\ \citenamefont {Kneschke}()}]{Andersen:2018nzq}%
  \BibitemOpen
  \bibfield  {author} {\bibinfo {author} {\bibfnamefont {J.~O.}\ \bibnamefont {Andersen}}\ and\ \bibinfo {author} {\bibfnamefont {P.}~\bibnamefont {Kneschke}},\ }\href@noop {} {\ }\Eprint {http://arxiv.org/abs/1807.08951} {arXiv:1807.08951 [hep-ph]} \BibitemShut {NoStop}%
\bibitem [{\citenamefont {Mannarelli}(2019)}]{Mannarelli:2019hgn}%
  \BibitemOpen
  \bibfield  {author} {\bibinfo {author} {\bibfnamefont {M.}~\bibnamefont {Mannarelli}},\ }\href {\doibase 10.3390/particles2030025} {\bibfield  {journal} {\bibinfo  {journal} {Particles}\ }\textbf {\bibinfo {volume} {2}},\ \bibinfo {pages} {411} (\bibinfo {year} {2019})},\ \Eprint {http://arxiv.org/abs/1908.02042} {arXiv:1908.02042 [hep-ph]} \BibitemShut {NoStop}%
\bibitem [{\citenamefont {Stashko}\ \emph {et~al.}(2023)\citenamefont {Stashko}, \citenamefont {Savchuk}, \citenamefont {Satarov}, \citenamefont {Mishustin}, \citenamefont {Gorenstein},\ and\ \citenamefont {Zhdanov}}]{Stashko:2023gnn}%
  \BibitemOpen
  \bibfield  {author} {\bibinfo {author} {\bibfnamefont {O.~S.}\ \bibnamefont {Stashko}}, \bibinfo {author} {\bibfnamefont {O.~V.}\ \bibnamefont {Savchuk}}, \bibinfo {author} {\bibfnamefont {L.~M.}\ \bibnamefont {Satarov}}, \bibinfo {author} {\bibfnamefont {I.~N.}\ \bibnamefont {Mishustin}}, \bibinfo {author} {\bibfnamefont {M.~I.}\ \bibnamefont {Gorenstein}}, \ and\ \bibinfo {author} {\bibfnamefont {V.~I.}\ \bibnamefont {Zhdanov}},\ }\href {\doibase 10.1103/PhysRevD.107.114025} {\bibfield  {journal} {\bibinfo  {journal} {Phys. Rev. D}\ }\textbf {\bibinfo {volume} {107}},\ \bibinfo {pages} {114025} (\bibinfo {year} {2023})},\ \Eprint {http://arxiv.org/abs/2303.06190} {arXiv:2303.06190 [hep-ph]} \BibitemShut {NoStop}%
\bibitem [{\citenamefont {Georgi}(1984)}]{Georgi:1985kw}%
  \BibitemOpen
  \bibfield  {author} {\bibinfo {author} {\bibfnamefont {H.}~\bibnamefont {Georgi}},\ }\href@noop {} {\emph {\bibinfo {title} {Weak Interactions and Modern Particle Theory}}}\ (\bibinfo  {publisher} {Benjamin/Cummings Publishing Company},\ \bibinfo {year} {1984})\BibitemShut {NoStop}%
\bibitem [{\citenamefont {Peskin}\ and\ \citenamefont {Schroeder}(1995)}]{peskin1995introduction}%
  \BibitemOpen
  \bibfield  {author} {\bibinfo {author} {\bibfnamefont {M.}~\bibnamefont {Peskin}}\ and\ \bibinfo {author} {\bibfnamefont {D.}~\bibnamefont {Schroeder}},\ }\href {https://books.google.it/books?id=i35LALN0GosC} {\emph {\bibinfo {title} {An Introduction to Quantum Field Theory}}},\ Advanced book classics\ (\bibinfo  {publisher} {Avalon Publishing},\ \bibinfo {year} {1995})\BibitemShut {NoStop}%
\bibitem [{\citenamefont {Prunier}\ \emph {et~al.}(2023)\citenamefont {Prunier}, \citenamefont {Morrás}, \citenamefont {Siles}, \citenamefont {Clesse}, \citenamefont {García-Bellido},\ and\ \citenamefont {Morales}}]{prunier2023analysis}%
  \BibitemOpen
  \bibfield  {author} {\bibinfo {author} {\bibfnamefont {M.}~\bibnamefont {Prunier}}, \bibinfo {author} {\bibfnamefont {G.}~\bibnamefont {Morrás}}, \bibinfo {author} {\bibfnamefont {J.~F.~N.}\ \bibnamefont {Siles}}, \bibinfo {author} {\bibfnamefont {S.}~\bibnamefont {Clesse}}, \bibinfo {author} {\bibfnamefont {J.}~\bibnamefont {García-Bellido}}, \ and\ \bibinfo {author} {\bibfnamefont {E.~R.}\ \bibnamefont {Morales}},\ }\href@noop {} {\enquote {\bibinfo {title} {Analysis of the subsolar-mass black hole candidate ssm200308 from the second part of the third observing run of advanced ligo-virgo},}\ } (\bibinfo {year} {2023}),\ \Eprint {http://arxiv.org/abs/2311.16085} {arXiv:2311.16085 [gr-qc]} \BibitemShut {NoStop}%
\bibitem [{\citenamefont {{Shapiro}}\ and\ \citenamefont {{Teukolsky}}(1983)}]{Shapiro-Teukolsky}%
  \BibitemOpen
  \bibfield  {author} {\bibinfo {author} {\bibfnamefont {S.~L.}\ \bibnamefont {{Shapiro}}}\ and\ \bibinfo {author} {\bibfnamefont {S.~A.}\ \bibnamefont {{Teukolsky}}},\ }\href@noop {} {\emph {\bibinfo {title} {Research supported by the National Science Foundation.~New York, Wiley-Interscience, 1983, 663 p.}}}\ (\bibinfo {year} {1983})\BibitemShut {NoStop}%
\bibitem [{\citenamefont {Douchin}\ and\ \citenamefont {Haensel}(2001)}]{Douchin:2001sv}%
  \BibitemOpen
  \bibfield  {author} {\bibinfo {author} {\bibfnamefont {F.}~\bibnamefont {Douchin}}\ and\ \bibinfo {author} {\bibfnamefont {P.}~\bibnamefont {Haensel}},\ }\href {\doibase 10.1051/0004-6361:20011402} {\bibfield  {journal} {\bibinfo  {journal} {Astron. Astrophys.}\ }\textbf {\bibinfo {volume} {380}},\ \bibinfo {pages} {151} (\bibinfo {year} {2001})},\ \Eprint {http://arxiv.org/abs/astro-ph/0111092} {arXiv:astro-ph/0111092 [astro-ph]} \BibitemShut {NoStop}%
\bibitem [{\citenamefont {Page}\ and\ \citenamefont {Reddy}(2006)}]{Page:2006ud}%
  \BibitemOpen
  \bibfield  {author} {\bibinfo {author} {\bibfnamefont {D.}~\bibnamefont {Page}}\ and\ \bibinfo {author} {\bibfnamefont {S.}~\bibnamefont {Reddy}},\ }\href {\doibase 10.1146/annurev.nucl.56.080805.140600} {\bibfield  {journal} {\bibinfo  {journal} {Ann. Rev. Nucl. Part. Sci.}\ }\textbf {\bibinfo {volume} {56}},\ \bibinfo {pages} {327} (\bibinfo {year} {2006})},\ \Eprint {http://arxiv.org/abs/astro-ph/0608360} {arXiv:astro-ph/0608360} \BibitemShut {NoStop}%
\bibitem [{\citenamefont {Blaschke}\ and\ \citenamefont {Chamel}(2018)}]{Blaschke:2018mqw}%
  \BibitemOpen
  \bibfield  {author} {\bibinfo {author} {\bibfnamefont {D.}~\bibnamefont {Blaschke}}\ and\ \bibinfo {author} {\bibfnamefont {N.}~\bibnamefont {Chamel}},\ }\href {\doibase 10.1007/978-3-319-97616-7_7} {\bibfield  {journal} {\bibinfo  {journal} {Astrophys. Space Sci. Libr.}\ }\textbf {\bibinfo {volume} {457}},\ \bibinfo {pages} {337} (\bibinfo {year} {2018})},\ \Eprint {http://arxiv.org/abs/1803.01836} {arXiv:1803.01836 [nucl-th]} \BibitemShut {NoStop}%
\bibitem [{\citenamefont {Lattimer}(2021)}]{Lattimer:2021emm}%
  \BibitemOpen
  \bibfield  {author} {\bibinfo {author} {\bibfnamefont {J.~M.}\ \bibnamefont {Lattimer}},\ }\href {\doibase 10.1146/annurev-nucl-102419-124827} {\bibfield  {journal} {\bibinfo  {journal} {Ann. Rev. Nucl. Part. Sci.}\ }\textbf {\bibinfo {volume} {71}},\ \bibinfo {pages} {433} (\bibinfo {year} {2021})}\BibitemShut {NoStop}%
\bibitem [{\citenamefont {Burgio}\ \emph {et~al.}(2021)\citenamefont {Burgio}, \citenamefont {Schulze}, \citenamefont {Vidana},\ and\ \citenamefont {Wei}}]{Burgio:2021vgk}%
  \BibitemOpen
  \bibfield  {author} {\bibinfo {author} {\bibfnamefont {G.~F.}\ \bibnamefont {Burgio}}, \bibinfo {author} {\bibfnamefont {H.~J.}\ \bibnamefont {Schulze}}, \bibinfo {author} {\bibfnamefont {I.}~\bibnamefont {Vidana}}, \ and\ \bibinfo {author} {\bibfnamefont {J.~B.}\ \bibnamefont {Wei}},\ }\href {\doibase 10.1016/j.ppnp.2021.103879} {\bibfield  {journal} {\bibinfo  {journal} {Prog. Part. Nucl. Phys.}\ }\textbf {\bibinfo {volume} {120}},\ \bibinfo {pages} {103879} (\bibinfo {year} {2021})},\ \Eprint {http://arxiv.org/abs/2105.03747} {arXiv:2105.03747 [nucl-th]} \BibitemShut {NoStop}%
\bibitem [{\citenamefont {{Colpi}}\ \emph {et~al.}(1989)\citenamefont {{Colpi}}, \citenamefont {{Shapiro}},\ and\ \citenamefont {{Teukolsky}}}]{1989ApJ...339..318C}%
  \BibitemOpen
  \bibfield  {author} {\bibinfo {author} {\bibfnamefont {M.}~\bibnamefont {{Colpi}}}, \bibinfo {author} {\bibfnamefont {S.~L.}\ \bibnamefont {{Shapiro}}}, \ and\ \bibinfo {author} {\bibfnamefont {S.~A.}\ \bibnamefont {{Teukolsky}}},\ }\href {\doibase 10.1086/167299} {\bibfield  {journal} {\bibinfo  {journal} {\apj}\ }\textbf {\bibinfo {volume} {339}},\ \bibinfo {pages} {318} (\bibinfo {year} {1989})}\BibitemShut {NoStop}%
\bibitem [{\citenamefont {Strobel}\ \emph {et~al.}(1999)\citenamefont {Strobel}, \citenamefont {Schaab},\ and\ \citenamefont {Weigel}}]{Strobel:1999vn}%
  \BibitemOpen
  \bibfield  {author} {\bibinfo {author} {\bibfnamefont {K.}~\bibnamefont {Strobel}}, \bibinfo {author} {\bibfnamefont {C.}~\bibnamefont {Schaab}}, \ and\ \bibinfo {author} {\bibfnamefont {M.~K.}\ \bibnamefont {Weigel}},\ }\href@noop {} {\bibfield  {journal} {\bibinfo  {journal} {Astron. Astrophys.}\ }\textbf {\bibinfo {volume} {350}},\ \bibinfo {pages} {497} (\bibinfo {year} {1999})},\ \Eprint {http://arxiv.org/abs/astro-ph/9908132} {arXiv:astro-ph/9908132 [astro-ph]} \BibitemShut {NoStop}%
\bibitem [{\citenamefont {Suwa}\ \emph {et~al.}(2018)\citenamefont {Suwa}, \citenamefont {Yoshida}, \citenamefont {Shibata}, \citenamefont {Umeda},\ and\ \citenamefont {Takahashi}}]{Suwa:2018uni}%
  \BibitemOpen
  \bibfield  {author} {\bibinfo {author} {\bibfnamefont {Y.}~\bibnamefont {Suwa}}, \bibinfo {author} {\bibfnamefont {T.}~\bibnamefont {Yoshida}}, \bibinfo {author} {\bibfnamefont {M.}~\bibnamefont {Shibata}}, \bibinfo {author} {\bibfnamefont {H.}~\bibnamefont {Umeda}}, \ and\ \bibinfo {author} {\bibfnamefont {K.}~\bibnamefont {Takahashi}},\ }\href {\doibase 10.1093/mnras/sty2460} {\bibfield  {journal} {\bibinfo  {journal} {Mon. Not. Roy. Astron. Soc.}\ }\textbf {\bibinfo {volume} {481}},\ \bibinfo {pages} {3305} (\bibinfo {year} {2018})},\ \Eprint {http://arxiv.org/abs/1808.02328} {arXiv:1808.02328 [astro-ph.HE]} \BibitemShut {NoStop}%
\bibitem [{\citenamefont {Kleihaus}\ \emph {et~al.}(2005)\citenamefont {Kleihaus}, \citenamefont {Kunz},\ and\ \citenamefont {List}}]{Kleihaus:2005me}%
  \BibitemOpen
  \bibfield  {author} {\bibinfo {author} {\bibfnamefont {B.}~\bibnamefont {Kleihaus}}, \bibinfo {author} {\bibfnamefont {J.}~\bibnamefont {Kunz}}, \ and\ \bibinfo {author} {\bibfnamefont {M.}~\bibnamefont {List}},\ }\href {\doibase 10.1103/PhysRevD.72.064002} {\bibfield  {journal} {\bibinfo  {journal} {Phys. Rev. D}\ }\textbf {\bibinfo {volume} {72}},\ \bibinfo {pages} {064002} (\bibinfo {year} {2005})},\ \Eprint {http://arxiv.org/abs/gr-qc/0505143} {arXiv:gr-qc/0505143} \BibitemShut {NoStop}%
\bibitem [{\citenamefont {Kleihaus}\ \emph {et~al.}(2008)\citenamefont {Kleihaus}, \citenamefont {Kunz}, \citenamefont {List},\ and\ \citenamefont {Schaffer}}]{Kleihaus:2007vk}%
  \BibitemOpen
  \bibfield  {author} {\bibinfo {author} {\bibfnamefont {B.}~\bibnamefont {Kleihaus}}, \bibinfo {author} {\bibfnamefont {J.}~\bibnamefont {Kunz}}, \bibinfo {author} {\bibfnamefont {M.}~\bibnamefont {List}}, \ and\ \bibinfo {author} {\bibfnamefont {I.}~\bibnamefont {Schaffer}},\ }\href {\doibase 10.1103/PhysRevD.77.064025} {\bibfield  {journal} {\bibinfo  {journal} {Phys. Rev. D}\ }\textbf {\bibinfo {volume} {77}},\ \bibinfo {pages} {064025} (\bibinfo {year} {2008})},\ \Eprint {http://arxiv.org/abs/0712.3742} {arXiv:0712.3742 [gr-qc]} \BibitemShut {NoStop}%
\bibitem [{\citenamefont {Khlopov}\ \emph {et~al.}(1985)\citenamefont {Khlopov}, \citenamefont {Malomed},\ and\ \citenamefont {Zeldovich}}]{Khlopov:1985}%
  \BibitemOpen
  \bibfield  {author} {\bibinfo {author} {\bibfnamefont {M.~Y.}\ \bibnamefont {Khlopov}}, \bibinfo {author} {\bibfnamefont {B.~A.}\ \bibnamefont {Malomed}}, \ and\ \bibinfo {author} {\bibfnamefont {Y.~B.}\ \bibnamefont {Zeldovich}},\ }\href {\doibase 10.1093/mnras/215.4.575} {\bibfield  {journal} {\bibinfo  {journal} {Monthly Notices of the Royal Astronomical Society}\ }\textbf {\bibinfo {volume} {215}},\ \bibinfo {pages} {575} (\bibinfo {year} {1985})},\ \Eprint {http://arxiv.org/abs/https://academic.oup.com/mnras/article-pdf/215/4/575/4082842/mnras215-0575.pdf} {https://academic.oup.com/mnras/article-pdf/215/4/575/4082842/mnras215-0575.pdf} \BibitemShut {NoStop}%
\bibitem [{\citenamefont {Colpi}\ \emph {et~al.}(1986)\citenamefont {Colpi}, \citenamefont {Shapiro},\ and\ \citenamefont {Wasserman}}]{Colpi:1986ye}%
  \BibitemOpen
  \bibfield  {author} {\bibinfo {author} {\bibfnamefont {M.}~\bibnamefont {Colpi}}, \bibinfo {author} {\bibfnamefont {S.~L.}\ \bibnamefont {Shapiro}}, \ and\ \bibinfo {author} {\bibfnamefont {I.}~\bibnamefont {Wasserman}},\ }\href {\doibase 10.1103/PhysRevLett.57.2485} {\bibfield  {journal} {\bibinfo  {journal} {Phys. Rev. Lett.}\ }\textbf {\bibinfo {volume} {57}},\ \bibinfo {pages} {2485} (\bibinfo {year} {1986})}\BibitemShut {NoStop}%
\bibitem [{\citenamefont {Son}(2002)}]{Son:2002zn}%
  \BibitemOpen
  \bibfield  {author} {\bibinfo {author} {\bibfnamefont {D.~T.}\ \bibnamefont {Son}},\ }\href@noop {} {\  (\bibinfo {year} {2002})},\ \Eprint {http://arxiv.org/abs/hep-ph/0204199} {arXiv:hep-ph/0204199 [hep-ph]} \BibitemShut {NoStop}%
\bibitem [{\citenamefont {Mannarelli}\ and\ \citenamefont {Manuel}(2008)}]{Mannarelli:2008jq}%
  \BibitemOpen
  \bibfield  {author} {\bibinfo {author} {\bibfnamefont {M.}~\bibnamefont {Mannarelli}}\ and\ \bibinfo {author} {\bibfnamefont {C.}~\bibnamefont {Manuel}},\ }\href {\doibase 10.1103/PhysRevD.77.103014} {\bibfield  {journal} {\bibinfo  {journal} {Phys. Rev. D}\ }\textbf {\bibinfo {volume} {77}},\ \bibinfo {pages} {103014} (\bibinfo {year} {2008})},\ \Eprint {http://arxiv.org/abs/0802.0321} {arXiv:0802.0321 [hep-ph]} \BibitemShut {NoStop}%
\bibitem [{\citenamefont {Dalfovo}\ \emph {et~al.}(1999)\citenamefont {Dalfovo}, \citenamefont {Giorgini}, \citenamefont {Pitaevskii},\ and\ \citenamefont {Stringari}}]{Dalfovo:1999zz}%
  \BibitemOpen
  \bibfield  {author} {\bibinfo {author} {\bibfnamefont {F.}~\bibnamefont {Dalfovo}}, \bibinfo {author} {\bibfnamefont {S.}~\bibnamefont {Giorgini}}, \bibinfo {author} {\bibfnamefont {L.~P.}\ \bibnamefont {Pitaevskii}}, \ and\ \bibinfo {author} {\bibfnamefont {S.}~\bibnamefont {Stringari}},\ }\href {\doibase 10.1103/RevModPhys.71.463} {\bibfield  {journal} {\bibinfo  {journal} {Rev. Mod. Phys.}\ }\textbf {\bibinfo {volume} {71}},\ \bibinfo {pages} {463} (\bibinfo {year} {1999})},\ \Eprint {http://arxiv.org/abs/cond-mat/9806038} {arXiv:cond-mat/9806038} \BibitemShut {NoStop}%
\bibitem [{\citenamefont {Pitz}\ and\ \citenamefont {Schaffner-Bielich}(2023)}]{Pitz:2023ejc}%
  \BibitemOpen
  \bibfield  {author} {\bibinfo {author} {\bibfnamefont {S.~L.}\ \bibnamefont {Pitz}}\ and\ \bibinfo {author} {\bibfnamefont {J.}~\bibnamefont {Schaffner-Bielich}},\ }\href {\doibase 10.1103/PhysRevD.108.103043} {\bibfield  {journal} {\bibinfo  {journal} {Phys. Rev. D}\ }\textbf {\bibinfo {volume} {108}},\ \bibinfo {pages} {103043} (\bibinfo {year} {2023})},\ \Eprint {http://arxiv.org/abs/2308.01254} {arXiv:2308.01254 [astro-ph.HE]} \BibitemShut {NoStop}%
\bibitem [{\citenamefont {Lattimer}\ and\ \citenamefont {Prakash}(2016)}]{Lattimer:2015nhk}%
  \BibitemOpen
  \bibfield  {author} {\bibinfo {author} {\bibfnamefont {J.~M.}\ \bibnamefont {Lattimer}}\ and\ \bibinfo {author} {\bibfnamefont {M.}~\bibnamefont {Prakash}},\ }\href {\doibase 10.1016/j.physrep.2015.12.005} {\bibfield  {journal} {\bibinfo  {journal} {Phys. Rept.}\ }\textbf {\bibinfo {volume} {621}},\ \bibinfo {pages} {127} (\bibinfo {year} {2016})},\ \Eprint {http://arxiv.org/abs/1512.07820} {arXiv:1512.07820 [astro-ph.SR]} \BibitemShut {NoStop}%
\bibitem [{\citenamefont {Rafiei~Karkevandi}\ \emph {et~al.}(2022)\citenamefont {Rafiei~Karkevandi}, \citenamefont {Shakeri}, \citenamefont {Sagun},\ and\ \citenamefont {Ivanytskyi}}]{PhysRevD.105.023001}%
  \BibitemOpen
  \bibfield  {author} {\bibinfo {author} {\bibfnamefont {D.}~\bibnamefont {Rafiei~Karkevandi}}, \bibinfo {author} {\bibfnamefont {S.}~\bibnamefont {Shakeri}}, \bibinfo {author} {\bibfnamefont {V.}~\bibnamefont {Sagun}}, \ and\ \bibinfo {author} {\bibfnamefont {O.}~\bibnamefont {Ivanytskyi}},\ }\href {\doibase 10.1103/PhysRevD.105.023001} {\bibfield  {journal} {\bibinfo  {journal} {Phys. Rev. D}\ }\textbf {\bibinfo {volume} {105}},\ \bibinfo {pages} {023001} (\bibinfo {year} {2022})}\BibitemShut {NoStop}%
\bibitem [{\citenamefont {{Tolman}}(1939)}]{Tolman}%
  \BibitemOpen
  \bibfield  {author} {\bibinfo {author} {\bibfnamefont {R.~C.}\ \bibnamefont {{Tolman}}},\ }\href {\doibase 10.1103/PhysRev.55.364} {\bibfield  {journal} {\bibinfo  {journal} {Phys. Rev.}\ }\textbf {\bibinfo {volume} {55}},\ \bibinfo {pages} {364} (\bibinfo {year} {1939})}\BibitemShut {NoStop}%
\bibitem [{\citenamefont {{Oppenheimer}}\ and\ \citenamefont {{Volkoff}}(1939)}]{Oppenheimer-Volkoff}%
  \BibitemOpen
  \bibfield  {author} {\bibinfo {author} {\bibfnamefont {J.~R.}\ \bibnamefont {{Oppenheimer}}}\ and\ \bibinfo {author} {\bibfnamefont {G.~M.}\ \bibnamefont {{Volkoff}}},\ }\href {\doibase 10.1103/PhysRev.55.374} {\bibfield  {journal} {\bibinfo  {journal} {Phys. Rev.}\ }\textbf {\bibinfo {volume} {55}},\ \bibinfo {pages} {374} (\bibinfo {year} {1939})}\BibitemShut {NoStop}%
\bibitem [{\citenamefont {Chabanat}\ \emph {et~al.}(1998)\citenamefont {Chabanat}, \citenamefont {Bonche}, \citenamefont {Haensel}, \citenamefont {Meyer},\ and\ \citenamefont {Schaeffer}}]{CHABANAT1998231}%
  \BibitemOpen
  \bibfield  {author} {\bibinfo {author} {\bibfnamefont {E.}~\bibnamefont {Chabanat}}, \bibinfo {author} {\bibfnamefont {P.}~\bibnamefont {Bonche}}, \bibinfo {author} {\bibfnamefont {P.}~\bibnamefont {Haensel}}, \bibinfo {author} {\bibfnamefont {J.}~\bibnamefont {Meyer}}, \ and\ \bibinfo {author} {\bibfnamefont {R.}~\bibnamefont {Schaeffer}},\ }\href {\doibase https://doi.org/10.1016/S0375-9474(98)00180-8} {\bibfield  {journal} {\bibinfo  {journal} {Nuclear Physics A}\ }\textbf {\bibinfo {volume} {635}},\ \bibinfo {pages} {231} (\bibinfo {year} {1998})}\BibitemShut {NoStop}%
\bibitem [{\citenamefont {{Bombaci, Ignazio}}\ and\ \citenamefont {{Logoteta, Domenico}}(2018)}]{BL}%
  \BibitemOpen
  \bibfield  {author} {\bibinfo {author} {\bibnamefont {{Bombaci, Ignazio}}}\ and\ \bibinfo {author} {\bibnamefont {{Logoteta, Domenico}}},\ }\href {\doibase 10.1051/0004-6361/201731604} {\bibfield  {journal} {\bibinfo  {journal} {A\&A}\ }\textbf {\bibinfo {volume} {609}},\ \bibinfo {pages} {A128} (\bibinfo {year} {2018})}\BibitemShut {NoStop}%
\bibitem [{\citenamefont {Grams}\ \emph {et~al.}(2022)\citenamefont {Grams}, \citenamefont {Margueron}, \citenamefont {Somasundaram},\ and\ \citenamefont {Reddy}}]{Grams2022}%
  \BibitemOpen
  \bibfield  {author} {\bibinfo {author} {\bibfnamefont {G.}~\bibnamefont {Grams}}, \bibinfo {author} {\bibfnamefont {J.}~\bibnamefont {Margueron}}, \bibinfo {author} {\bibfnamefont {R.}~\bibnamefont {Somasundaram}}, \ and\ \bibinfo {author} {\bibfnamefont {S.}~\bibnamefont {Reddy}},\ }\href {\doibase 10.1140/epja/s10050-022-00706-w} {\bibfield  {journal} {\bibinfo  {journal} {The European Physical Journal A}\ }\textbf {\bibinfo {volume} {58}},\ \bibinfo {pages} {56} (\bibinfo {year} {2022})}\BibitemShut {NoStop}%
\bibitem [{\citenamefont {Akmal}\ \emph {et~al.}(1998)\citenamefont {Akmal}, \citenamefont {Pandharipande},\ and\ \citenamefont {Ravenhall}}]{PhysRevC.58.1804}%
  \BibitemOpen
  \bibfield  {author} {\bibinfo {author} {\bibfnamefont {A.}~\bibnamefont {Akmal}}, \bibinfo {author} {\bibfnamefont {V.~R.}\ \bibnamefont {Pandharipande}}, \ and\ \bibinfo {author} {\bibfnamefont {D.~G.}\ \bibnamefont {Ravenhall}},\ }\href {\doibase 10.1103/PhysRevC.58.1804} {\bibfield  {journal} {\bibinfo  {journal} {Phys. Rev. C}\ }\textbf {\bibinfo {volume} {58}},\ \bibinfo {pages} {1804} (\bibinfo {year} {1998})}\BibitemShut {NoStop}%
\bibitem [{\citenamefont {Typel}\ \emph {et~al.}(2010)\citenamefont {Typel}, \citenamefont {R\"opke}, \citenamefont {Kl\"ahn}, \citenamefont {Blaschke},\ and\ \citenamefont {Wolter}}]{PhysRevC.81.015803}%
  \BibitemOpen
  \bibfield  {author} {\bibinfo {author} {\bibfnamefont {S.}~\bibnamefont {Typel}}, \bibinfo {author} {\bibfnamefont {G.}~\bibnamefont {R\"opke}}, \bibinfo {author} {\bibfnamefont {T.}~\bibnamefont {Kl\"ahn}}, \bibinfo {author} {\bibfnamefont {D.}~\bibnamefont {Blaschke}}, \ and\ \bibinfo {author} {\bibfnamefont {H.~H.}\ \bibnamefont {Wolter}},\ }\href {\doibase 10.1103/PhysRevC.81.015803} {\bibfield  {journal} {\bibinfo  {journal} {Phys. Rev. C}\ }\textbf {\bibinfo {volume} {81}},\ \bibinfo {pages} {015803} (\bibinfo {year} {2010})}\BibitemShut {NoStop}%
\bibitem [{\citenamefont {Romani}\ \emph {et~al.}(2021)\citenamefont {Romani}, \citenamefont {Kandel}, \citenamefont {Filippenko}, \citenamefont {Brink},\ and\ \citenamefont {Zheng}}]{Romani_2021}%
  \BibitemOpen
  \bibfield  {author} {\bibinfo {author} {\bibfnamefont {R.~W.}\ \bibnamefont {Romani}}, \bibinfo {author} {\bibfnamefont {D.}~\bibnamefont {Kandel}}, \bibinfo {author} {\bibfnamefont {A.~V.}\ \bibnamefont {Filippenko}}, \bibinfo {author} {\bibfnamefont {T.~G.}\ \bibnamefont {Brink}}, \ and\ \bibinfo {author} {\bibfnamefont {W.}~\bibnamefont {Zheng}},\ }\href {\doibase 10.3847/2041-8213/abe2b4} {\bibfield  {journal} {\bibinfo  {journal} {The Astrophysical Journal Letters}\ }\textbf {\bibinfo {volume} {908}},\ \bibinfo {pages} {L46} (\bibinfo {year} {2021})}\BibitemShut {NoStop}%
\bibitem [{\citenamefont {Antoniadis}\ \emph {et~al.}(2013)\citenamefont {Antoniadis} \emph {et~al.}}]{Antoniadis_2013}%
  \BibitemOpen
  \bibfield  {author} {\bibinfo {author} {\bibfnamefont {J.}~\bibnamefont {Antoniadis}} \emph {et~al.},\ }\href {\doibase 10.1126/science.1233232} {\bibfield  {journal} {\bibinfo  {journal} {Science}\ }\textbf {\bibinfo {volume} {340}} (\bibinfo {year} {2013}),\ 10.1126/science.1233232}\BibitemShut {NoStop}%
\bibitem [{\citenamefont {Pani}\ and\ \citenamefont {Franciolini}(tion)}]{Pani}%
  \BibitemOpen
  \bibfield  {author} {\bibinfo {author} {\bibfnamefont {P.}~\bibnamefont {Pani}}\ and\ \bibinfo {author} {\bibfnamefont {G.}~\bibnamefont {Franciolini}},\ }\href@noop {} {} (\bibinfo {year} {Private communication})\BibitemShut {NoStop}%
\bibitem [{\citenamefont {LIGO}()}]{LIGO-sens}%
  \BibitemOpen
  \bibfield  {author} {\bibinfo {author} {\bibnamefont {LIGO}},\ }\href {https://dcc.ligo.org/LIGO-T1500293/public} {\enquote {\bibinfo {title} {Public ligo document t1500293-v13},}\ }\BibitemShut {NoStop}%
\bibitem [{\citenamefont {Crescimbeni}\ \emph {et~al.}(2024)\citenamefont {Crescimbeni}, \citenamefont {Franciolini}, \citenamefont {Pani},\ and\ \citenamefont {Riotto}}]{crescimbeni2024primordial}%
  \BibitemOpen
  \bibfield  {author} {\bibinfo {author} {\bibfnamefont {F.}~\bibnamefont {Crescimbeni}}, \bibinfo {author} {\bibfnamefont {G.}~\bibnamefont {Franciolini}}, \bibinfo {author} {\bibfnamefont {P.}~\bibnamefont {Pani}}, \ and\ \bibinfo {author} {\bibfnamefont {A.}~\bibnamefont {Riotto}},\ }\href@noop {} {\enquote {\bibinfo {title} {Primordial black holes or else? tidal tests on subsolar mass gravitational-wave observations},}\ } (\bibinfo {year} {2024}),\ \Eprint {http://arxiv.org/abs/2402.18656} {arXiv:2402.18656 [astro-ph.HE]} \BibitemShut {NoStop}%
\bibitem [{\citenamefont {Branchesi}\ \emph {et~al.}(2023)\citenamefont {Branchesi} \emph {et~al.}}]{Branchesi:2023mws}%
  \BibitemOpen
  \bibfield  {author} {\bibinfo {author} {\bibfnamefont {M.}~\bibnamefont {Branchesi}} \emph {et~al.},\ }\href {\doibase 10.1088/1475-7516/2023/07/068} {\bibfield  {journal} {\bibinfo  {journal} {JCAP}\ }\textbf {\bibinfo {volume} {07}},\ \bibinfo {pages} {068} (\bibinfo {year} {2023})},\ \Eprint {http://arxiv.org/abs/2303.15923} {arXiv:2303.15923 [gr-qc]} \BibitemShut {NoStop}%
\bibitem [{\citenamefont {Riley}\ \emph {et~al.}(2019)\citenamefont {Riley} \emph {et~al.}}]{Riley:2019yda}%
  \BibitemOpen
  \bibfield  {author} {\bibinfo {author} {\bibfnamefont {T.~E.}\ \bibnamefont {Riley}} \emph {et~al.},\ }\href {\doibase 10.3847/2041-8213/ab481c} {\bibfield  {journal} {\bibinfo  {journal} {Astrophys. J. Lett.}\ }\textbf {\bibinfo {volume} {887}},\ \bibinfo {pages} {L21} (\bibinfo {year} {2019})},\ \Eprint {http://arxiv.org/abs/1912.05702} {arXiv:1912.05702 [astro-ph.HE]} \BibitemShut {NoStop}%
\bibitem [{\citenamefont {Abbott}\ \emph {et~al.}(2017{\natexlab{a}})\citenamefont {Abbott} \emph {et~al.}}]{TheLIGOScientific:2017qsa}%
  \BibitemOpen
  \bibfield  {author} {\bibinfo {author} {\bibfnamefont {B.}~\bibnamefont {Abbott}} \emph {et~al.} (\bibinfo {collaboration} {Virgo, LIGO Scientific}),\ }\href {\doibase 10.1103/PhysRevLett.119.161101} {\bibfield  {journal} {\bibinfo  {journal} {Phys. Rev. Lett.}\ }\textbf {\bibinfo {volume} {119}},\ \bibinfo {pages} {161101} (\bibinfo {year} {2017}{\natexlab{a}})},\ \Eprint {http://arxiv.org/abs/1710.05832} {arXiv:1710.05832 [gr-qc]} \BibitemShut {NoStop}%
\bibitem [{\citenamefont {Abbott}\ \emph {et~al.}(2017{\natexlab{b}})\citenamefont {Abbott} \emph {et~al.}}]{LIGOScientific:2017ync}%
  \BibitemOpen
  \bibfield  {author} {\bibinfo {author} {\bibfnamefont {B.~P.}\ \bibnamefont {Abbott}} \emph {et~al.} (\bibinfo {collaboration} {LIGO Scientific, Virgo, Fermi GBM, INTEGRAL, IceCube, AstroSat Cadmium Zinc Telluride Imager Team, IPN, Insight-Hxmt, ANTARES, Swift, AGILE Team, 1M2H Team, Dark Energy Camera GW-EM, DES, DLT40, GRAWITA, Fermi-LAT, ATCA, ASKAP, Las Cumbres Observatory Group, OzGrav, DWF (Deeper Wider Faster Program), AST3, CAASTRO, VINROUGE, MASTER, J-GEM, GROWTH, JAGWAR, CaltechNRAO, TTU-NRAO, NuSTAR, Pan-STARRS, MAXI Team, TZAC Consortium, KU, Nordic Optical Telescope, ePESSTO, GROND, Texas Tech University, SALT Group, TOROS, BOOTES, MWA, CALET, IKI-GW Follow-up, H.E.S.S., LOFAR, LWA, HAWC, Pierre Auger, ALMA, Euro VLBI Team, Pi of Sky, Chandra Team at McGill University, DFN, ATLAS Telescopes, High Time Resolution Universe Survey, RIMAS, RATIR, SKA South Africa/MeerKAT}),\ }\href {\doibase 10.3847/2041-8213/aa91c9} {\bibfield  {journal} {\bibinfo  {journal} {Astrophys. J. Lett.}\ }\textbf
  {\bibinfo {volume} {848}},\ \bibinfo {pages} {L12} (\bibinfo {year} {2017}{\natexlab{b}})},\ \Eprint {http://arxiv.org/abs/1710.05833} {arXiv:1710.05833 [astro-ph.HE]} \BibitemShut {NoStop}%
\bibitem [{\citenamefont {Hinderer}(2008)}]{Hinderer:2007mb}%
  \BibitemOpen
  \bibfield  {author} {\bibinfo {author} {\bibfnamefont {T.}~\bibnamefont {Hinderer}},\ }\href {\doibase 10.1086/533487} {\bibfield  {journal} {\bibinfo  {journal} {Astrophys. J.}\ }\textbf {\bibinfo {volume} {677}},\ \bibinfo {pages} {1216} (\bibinfo {year} {2008})},\ \Eprint {http://arxiv.org/abs/0711.2420} {arXiv:0711.2420 [astro-ph]} \BibitemShut {NoStop}%
\bibitem [{\citenamefont {Flanagan}\ and\ \citenamefont {Hinderer}(2008)}]{Flanagan:2007ix}%
  \BibitemOpen
  \bibfield  {author} {\bibinfo {author} {\bibfnamefont {E.~E.}\ \bibnamefont {Flanagan}}\ and\ \bibinfo {author} {\bibfnamefont {T.}~\bibnamefont {Hinderer}},\ }\href {\doibase 10.1103/PhysRevD.77.021502} {\bibfield  {journal} {\bibinfo  {journal} {Phys. Rev.}\ }\textbf {\bibinfo {volume} {D77}},\ \bibinfo {pages} {021502} (\bibinfo {year} {2008})},\ \Eprint {http://arxiv.org/abs/0709.1915} {arXiv:0709.1915 [astro-ph]} \BibitemShut {NoStop}%
\bibitem [{\citenamefont {Damour}\ and\ \citenamefont {Nagar}(2009)}]{Damour:2009vw}%
  \BibitemOpen
  \bibfield  {author} {\bibinfo {author} {\bibfnamefont {T.}~\bibnamefont {Damour}}\ and\ \bibinfo {author} {\bibfnamefont {A.}~\bibnamefont {Nagar}},\ }\href {\doibase 10.1103/PhysRevD.80.084035} {\bibfield  {journal} {\bibinfo  {journal} {Phys. Rev. D}\ }\textbf {\bibinfo {volume} {80}},\ \bibinfo {pages} {084035} (\bibinfo {year} {2009})},\ \Eprint {http://arxiv.org/abs/0906.0096} {arXiv:0906.0096 [gr-qc]} \BibitemShut {NoStop}%
\bibitem [{\citenamefont {Postnikov}\ \emph {et~al.}(2010)\citenamefont {Postnikov}, \citenamefont {Prakash},\ and\ \citenamefont {Lattimer}}]{Postnikov:2010yn}%
  \BibitemOpen
  \bibfield  {author} {\bibinfo {author} {\bibfnamefont {S.}~\bibnamefont {Postnikov}}, \bibinfo {author} {\bibfnamefont {M.}~\bibnamefont {Prakash}}, \ and\ \bibinfo {author} {\bibfnamefont {J.~M.}\ \bibnamefont {Lattimer}},\ }\href {\doibase 10.1103/PhysRevD.82.024016} {\bibfield  {journal} {\bibinfo  {journal} {Phys. Rev.}\ }\textbf {\bibinfo {volume} {D82}},\ \bibinfo {pages} {024016} (\bibinfo {year} {2010})},\ \Eprint {http://arxiv.org/abs/1004.5098} {arXiv:1004.5098 [astro-ph.SR]} \BibitemShut {NoStop}%
\bibitem [{\citenamefont {Bertone}\ \emph {et~al.}(2020)\citenamefont {Bertone} \emph {et~al.}}]{Bertone:2019irm}%
  \BibitemOpen
  \bibfield  {author} {\bibinfo {author} {\bibfnamefont {G.}~\bibnamefont {Bertone}} \emph {et~al.},\ }\href {\doibase 10.21468/SciPostPhysCore.3.2.007} {\bibfield  {journal} {\bibinfo  {journal} {SciPost Phys. Core}\ }\textbf {\bibinfo {volume} {3}},\ \bibinfo {pages} {007} (\bibinfo {year} {2020})},\ \Eprint {http://arxiv.org/abs/1907.10610} {arXiv:1907.10610 [astro-ph.CO]} \BibitemShut {NoStop}%
\bibitem [{\citenamefont {Rosa}\ and\ \citenamefont {Rubiera-Garcia}(2022)}]{Rosa:2022tfv}%
  \BibitemOpen
  \bibfield  {author} {\bibinfo {author} {\bibfnamefont {J.~a.~L.}\ \bibnamefont {Rosa}}\ and\ \bibinfo {author} {\bibfnamefont {D.}~\bibnamefont {Rubiera-Garcia}},\ }\href {\doibase 10.1103/PhysRevD.106.084004} {\bibfield  {journal} {\bibinfo  {journal} {Phys. Rev. D}\ }\textbf {\bibinfo {volume} {106}},\ \bibinfo {pages} {084004} (\bibinfo {year} {2022})},\ \Eprint {http://arxiv.org/abs/2204.12949} {arXiv:2204.12949 [gr-qc]} \BibitemShut {NoStop}%
\bibitem [{\citenamefont {Palenzuela}\ \emph {et~al.}(2008)\citenamefont {Palenzuela}, \citenamefont {Lehner},\ and\ \citenamefont {Liebling}}]{Palenzuela:2007dm}%
  \BibitemOpen
  \bibfield  {author} {\bibinfo {author} {\bibfnamefont {C.}~\bibnamefont {Palenzuela}}, \bibinfo {author} {\bibfnamefont {L.}~\bibnamefont {Lehner}}, \ and\ \bibinfo {author} {\bibfnamefont {S.~L.}\ \bibnamefont {Liebling}},\ }\href {\doibase 10.1103/PhysRevD.77.044036} {\bibfield  {journal} {\bibinfo  {journal} {Phys. Rev. D}\ }\textbf {\bibinfo {volume} {77}},\ \bibinfo {pages} {044036} (\bibinfo {year} {2008})},\ \Eprint {http://arxiv.org/abs/0706.2435} {arXiv:0706.2435 [gr-qc]} \BibitemShut {NoStop}%
\bibitem [{\citenamefont {Bezares}\ and\ \citenamefont {Palenzuela}(2018)}]{Bezares:2018qwa}%
  \BibitemOpen
  \bibfield  {author} {\bibinfo {author} {\bibfnamefont {M.}~\bibnamefont {Bezares}}\ and\ \bibinfo {author} {\bibfnamefont {C.}~\bibnamefont {Palenzuela}},\ }\href {\doibase 10.1088/1361-6382/aae87c} {\bibfield  {journal} {\bibinfo  {journal} {Class. Quant. Grav.}\ }\textbf {\bibinfo {volume} {35}},\ \bibinfo {pages} {234002} (\bibinfo {year} {2018})},\ \Eprint {http://arxiv.org/abs/1808.10732} {arXiv:1808.10732 [gr-qc]} \BibitemShut {NoStop}%
\bibitem [{\citenamefont {Palenzuela}\ \emph {et~al.}(2017)\citenamefont {Palenzuela}, \citenamefont {Pani}, \citenamefont {Bezares}, \citenamefont {Cardoso}, \citenamefont {Lehner},\ and\ \citenamefont {Liebling}}]{PhysRevD.96.104058}%
  \BibitemOpen
  \bibfield  {author} {\bibinfo {author} {\bibfnamefont {C.}~\bibnamefont {Palenzuela}}, \bibinfo {author} {\bibfnamefont {P.}~\bibnamefont {Pani}}, \bibinfo {author} {\bibfnamefont {M.}~\bibnamefont {Bezares}}, \bibinfo {author} {\bibfnamefont {V.}~\bibnamefont {Cardoso}}, \bibinfo {author} {\bibfnamefont {L.}~\bibnamefont {Lehner}}, \ and\ \bibinfo {author} {\bibfnamefont {S.}~\bibnamefont {Liebling}},\ }\href {\doibase 10.1103/PhysRevD.96.104058} {\bibfield  {journal} {\bibinfo  {journal} {Phys. Rev. D}\ }\textbf {\bibinfo {volume} {96}},\ \bibinfo {pages} {104058} (\bibinfo {year} {2017})}\BibitemShut {NoStop}%
\bibitem [{\citenamefont {De~Pietri}\ \emph {et~al.}(2020)\citenamefont {De~Pietri}, \citenamefont {Feo}, \citenamefont {Font}, \citenamefont {L\"offler}, \citenamefont {Pasquali},\ and\ \citenamefont {Stergioulas}}]{DePietri:2019mti}%
  \BibitemOpen
  \bibfield  {author} {\bibinfo {author} {\bibfnamefont {R.}~\bibnamefont {De~Pietri}}, \bibinfo {author} {\bibfnamefont {A.}~\bibnamefont {Feo}}, \bibinfo {author} {\bibfnamefont {J.~A.}\ \bibnamefont {Font}}, \bibinfo {author} {\bibfnamefont {F.}~\bibnamefont {L\"offler}}, \bibinfo {author} {\bibfnamefont {M.}~\bibnamefont {Pasquali}}, \ and\ \bibinfo {author} {\bibfnamefont {N.}~\bibnamefont {Stergioulas}},\ }\href {\doibase 10.1103/PhysRevD.101.064052} {\bibfield  {journal} {\bibinfo  {journal} {Phys. Rev. D}\ }\textbf {\bibinfo {volume} {101}},\ \bibinfo {pages} {064052} (\bibinfo {year} {2020})},\ \Eprint {http://arxiv.org/abs/1910.04036} {arXiv:1910.04036 [gr-qc]} \BibitemShut {NoStop}%
\bibitem [{lor(1999)}]{loreneWS}%
  \BibitemOpen
  \href@noop {} {\enquote {\bibinfo {title} {\textsc{LORENE}: \textsc{L}angage \textsc{O}bjet pour la \textsc{RE}lativité \textsc{N}umériqu\textsc{E}},}\ }\bibinfo {howpublished} {\url{https://lorene.obspm.fr/}} (\bibinfo {year} {1999})\BibitemShut {NoStop}%
\bibitem [{\citenamefont {Gourgoulhon}\ \emph {et~al.}(2001)\citenamefont {Gourgoulhon}, \citenamefont {Grandclément}, \citenamefont {Taniguchi}, \citenamefont {Marck},\ and\ \citenamefont {Bonazzola}}]{Gourgoulhon_2001}%
  \BibitemOpen
  \bibfield  {author} {\bibinfo {author} {\bibfnamefont {E.}~\bibnamefont {Gourgoulhon}}, \bibinfo {author} {\bibfnamefont {P.}~\bibnamefont {Grandclément}}, \bibinfo {author} {\bibfnamefont {K.}~\bibnamefont {Taniguchi}}, \bibinfo {author} {\bibfnamefont {J.-A.}\ \bibnamefont {Marck}}, \ and\ \bibinfo {author} {\bibfnamefont {S.}~\bibnamefont {Bonazzola}},\ }\href {\doibase 10.1103/physrevd.63.064029} {\bibfield  {journal} {\bibinfo  {journal} {Physical Review D}\ }\textbf {\bibinfo {volume} {63}} (\bibinfo {year} {2001}),\ 10.1103/physrevd.63.064029}\BibitemShut {NoStop}%
\bibitem [{\citenamefont {Löffler}\ \emph {et~al.}(2012)\citenamefont {Löffler}, \citenamefont {Faber}, \citenamefont {Bentivegna}, \citenamefont {Bode}, \citenamefont {Diener}, \citenamefont {Haas}, \citenamefont {Hinder}, \citenamefont {Mundim}, \citenamefont {Ott}, \citenamefont {Schnetter}, \citenamefont {Allen}, \citenamefont {Campanelli},\ and\ \citenamefont {Laguna}}]{L_ffler_2012}%
  \BibitemOpen
  \bibfield  {author} {\bibinfo {author} {\bibfnamefont {F.}~\bibnamefont {Löffler}}, \bibinfo {author} {\bibfnamefont {J.}~\bibnamefont {Faber}}, \bibinfo {author} {\bibfnamefont {E.}~\bibnamefont {Bentivegna}}, \bibinfo {author} {\bibfnamefont {T.}~\bibnamefont {Bode}}, \bibinfo {author} {\bibfnamefont {P.}~\bibnamefont {Diener}}, \bibinfo {author} {\bibfnamefont {R.}~\bibnamefont {Haas}}, \bibinfo {author} {\bibfnamefont {I.}~\bibnamefont {Hinder}}, \bibinfo {author} {\bibfnamefont {B.~C.}\ \bibnamefont {Mundim}}, \bibinfo {author} {\bibfnamefont {C.~D.}\ \bibnamefont {Ott}}, \bibinfo {author} {\bibfnamefont {E.}~\bibnamefont {Schnetter}}, \bibinfo {author} {\bibfnamefont {G.}~\bibnamefont {Allen}}, \bibinfo {author} {\bibfnamefont {M.}~\bibnamefont {Campanelli}}, \ and\ \bibinfo {author} {\bibfnamefont {P.}~\bibnamefont {Laguna}},\ }\href {\doibase 10.1088/0264-9381/29/11/115001} {\bibfield  {journal} {\bibinfo  {journal} {Classical and Quantum Gravity}\ }\textbf {\bibinfo {volume} {29}},\ \bibinfo
  {pages} {115001} (\bibinfo {year} {2012})}\BibitemShut {NoStop}%
\bibitem [{\citenamefont {Hinder}\ \emph {et~al.}(2013)\citenamefont {Hinder} \emph {et~al.}}]{Hinder_2013}%
  \BibitemOpen
  \bibfield  {author} {\bibinfo {author} {\bibfnamefont {I.}~\bibnamefont {Hinder}} \emph {et~al.},\ }\href {\doibase 10.1088/0264-9381/31/2/025012} {\bibfield  {journal} {\bibinfo  {journal} {Classical and Quantum Gravity}\ }\textbf {\bibinfo {volume} {31}},\ \bibinfo {pages} {025012} (\bibinfo {year} {2013})}\BibitemShut {NoStop}%
\bibitem [{\citenamefont {Stergioulas}\ \emph {et~al.}(2011)\citenamefont {Stergioulas}, \citenamefont {Bauswein}, \citenamefont {Zagkouris},\ and\ \citenamefont {Janka}}]{Stergioulas:2011}%
  \BibitemOpen
  \bibfield  {author} {\bibinfo {author} {\bibfnamefont {N.}~\bibnamefont {Stergioulas}}, \bibinfo {author} {\bibfnamefont {A.}~\bibnamefont {Bauswein}}, \bibinfo {author} {\bibfnamefont {K.}~\bibnamefont {Zagkouris}}, \ and\ \bibinfo {author} {\bibfnamefont {H.-T.}\ \bibnamefont {Janka}},\ }\href {\doibase 10.1111/j.1365-2966.2011.19493.x} {\bibfield  {journal} {\bibinfo  {journal} {Monthly Notices of the Royal Astronomical Society}\ }\textbf {\bibinfo {volume} {418}},\ \bibinfo {pages} {427} (\bibinfo {year} {2011})},\ \Eprint {http://arxiv.org/abs/https://academic.oup.com/mnras/article-pdf/418/1/427/2849833/mnras0418-0427.pdf} {https://academic.oup.com/mnras/article-pdf/418/1/427/2849833/mnras0418-0427.pdf} \BibitemShut {NoStop}%
\bibitem [{\citenamefont {Kyutoku}\ \emph {et~al.}(2014)\citenamefont {Kyutoku}, \citenamefont {Shibata},\ and\ \citenamefont {Taniguchi}}]{PhysRevD.90.064006}%
  \BibitemOpen
  \bibfield  {author} {\bibinfo {author} {\bibfnamefont {K.}~\bibnamefont {Kyutoku}}, \bibinfo {author} {\bibfnamefont {M.}~\bibnamefont {Shibata}}, \ and\ \bibinfo {author} {\bibfnamefont {K.}~\bibnamefont {Taniguchi}},\ }\href {\doibase 10.1103/PhysRevD.90.064006} {\bibfield  {journal} {\bibinfo  {journal} {Phys. Rev. D}\ }\textbf {\bibinfo {volume} {90}},\ \bibinfo {pages} {064006} (\bibinfo {year} {2014})}\BibitemShut {NoStop}%
\bibitem [{\citenamefont {Dietrich}\ \emph {et~al.}(2015)\citenamefont {Dietrich}, \citenamefont {Moldenhauer}, \citenamefont {Johnson-McDaniel}, \citenamefont {Bernuzzi}, \citenamefont {Markakis}, \citenamefont {Brügmann},\ and\ \citenamefont {Tichy}}]{Dietrich_2015}%
  \BibitemOpen
  \bibfield  {author} {\bibinfo {author} {\bibfnamefont {T.}~\bibnamefont {Dietrich}}, \bibinfo {author} {\bibfnamefont {N.}~\bibnamefont {Moldenhauer}}, \bibinfo {author} {\bibfnamefont {N.~K.}\ \bibnamefont {Johnson-McDaniel}}, \bibinfo {author} {\bibfnamefont {S.}~\bibnamefont {Bernuzzi}}, \bibinfo {author} {\bibfnamefont {C.~M.}\ \bibnamefont {Markakis}}, \bibinfo {author} {\bibfnamefont {B.}~\bibnamefont {Brügmann}}, \ and\ \bibinfo {author} {\bibfnamefont {W.}~\bibnamefont {Tichy}},\ }\href {\doibase 10.1103/physrevd.92.124007} {\bibfield  {journal} {\bibinfo  {journal} {Physical Review D}\ }\textbf {\bibinfo {volume} {92}} (\bibinfo {year} {2015}),\ 10.1103/physrevd.92.124007}\BibitemShut {NoStop}%
\bibitem [{\citenamefont {Abbott}\ \emph {et~al.}(2020)\citenamefont {Abbott} \emph {et~al.}}]{LVK}%
  \BibitemOpen
  \bibfield  {author} {\bibinfo {author} {\bibfnamefont {B.}~\bibnamefont {Abbott}} \emph {et~al.},\ }\href {\doibase 10.1007/s41114-020-00026-9} {\bibfield  {journal} {\bibinfo  {journal} {Living Reviews in Relativity}\ }\textbf {\bibinfo {volume} {23}},\ \bibinfo {pages} {3} (\bibinfo {year} {2020})}\BibitemShut {NoStop}%
\bibitem [{\citenamefont {Bauswein}\ \emph {et~al.}(2016)\citenamefont {Bauswein}, \citenamefont {Stergioulas},\ and\ \citenamefont {Janka}}]{Bauswein2016}%
  \BibitemOpen
  \bibfield  {author} {\bibinfo {author} {\bibfnamefont {A.}~\bibnamefont {Bauswein}}, \bibinfo {author} {\bibfnamefont {N.}~\bibnamefont {Stergioulas}}, \ and\ \bibinfo {author} {\bibfnamefont {H.-T.}\ \bibnamefont {Janka}},\ }\href {\doibase 10.1140/epja/i2016-16056-7} {\bibfield  {journal} {\bibinfo  {journal} {The European Physical Journal A}\ }\textbf {\bibinfo {volume} {52}},\ \bibinfo {pages} {56} (\bibinfo {year} {2016})}\BibitemShut {NoStop}%
\bibitem [{\citenamefont {Bauswein}\ \emph {et~al.}(2012)\citenamefont {Bauswein}, \citenamefont {Janka}, \citenamefont {Hebeler},\ and\ \citenamefont {Schwenk}}]{PhysRevD.86.063001}%
  \BibitemOpen
  \bibfield  {author} {\bibinfo {author} {\bibfnamefont {A.}~\bibnamefont {Bauswein}}, \bibinfo {author} {\bibfnamefont {H.-T.}\ \bibnamefont {Janka}}, \bibinfo {author} {\bibfnamefont {K.}~\bibnamefont {Hebeler}}, \ and\ \bibinfo {author} {\bibfnamefont {A.}~\bibnamefont {Schwenk}},\ }\href {\doibase 10.1103/PhysRevD.86.063001} {\bibfield  {journal} {\bibinfo  {journal} {Phys. Rev. D}\ }\textbf {\bibinfo {volume} {86}},\ \bibinfo {pages} {063001} (\bibinfo {year} {2012})}\BibitemShut {NoStop}%
\bibitem [{\citenamefont {Franciolini}\ and\ \citenamefont {Pani}()}]{Pani:private}%
  \BibitemOpen
  \bibfield  {author} {\bibinfo {author} {\bibfnamefont {G.}~\bibnamefont {Franciolini}}\ and\ \bibinfo {author} {\bibfnamefont {P.}~\bibnamefont {Pani}},\ }\href@noop {} {\ }\Eprint {http://arxiv.org/abs/Private communication} {Private communication} \BibitemShut {NoStop}%
\bibitem [{\citenamefont {Rezzolla}\ and\ \citenamefont {Takami}(2016)}]{Rezzolla:2016nxn}%
  \BibitemOpen
  \bibfield  {author} {\bibinfo {author} {\bibfnamefont {L.}~\bibnamefont {Rezzolla}}\ and\ \bibinfo {author} {\bibfnamefont {K.}~\bibnamefont {Takami}},\ }\href {\doibase 10.1103/PhysRevD.93.124051} {\bibfield  {journal} {\bibinfo  {journal} {Phys. Rev. D}\ }\textbf {\bibinfo {volume} {93}},\ \bibinfo {pages} {124051} (\bibinfo {year} {2016})},\ \Eprint {http://arxiv.org/abs/1604.00246} {arXiv:1604.00246 [gr-qc]} \BibitemShut {NoStop}%
\bibitem [{\citenamefont {{Banyuls}}\ \emph {et~al.}(1997)\citenamefont {{Banyuls}}, \citenamefont {{Font}}, \citenamefont {{Ib{\'a}{\~n}ez}}, \citenamefont {{Mart{\'\i}}},\ and\ \citenamefont {{Miralles}}}]{1997ApJ...476..221B}%
  \BibitemOpen
  \bibfield  {author} {\bibinfo {author} {\bibfnamefont {F.}~\bibnamefont {{Banyuls}}}, \bibinfo {author} {\bibfnamefont {J.~A.}\ \bibnamefont {{Font}}}, \bibinfo {author} {\bibfnamefont {J.~M.}\ \bibnamefont {{Ib{\'a}{\~n}ez}}}, \bibinfo {author} {\bibfnamefont {J.~M.}\ \bibnamefont {{Mart{\'\i}}}}, \ and\ \bibinfo {author} {\bibfnamefont {J.~A.}\ \bibnamefont {{Miralles}}},\ }\href {\doibase 10.1086/303604} {\bibfield  {journal} {\bibinfo  {journal} {\apj}\ }\textbf {\bibinfo {volume} {476}},\ \bibinfo {pages} {221} (\bibinfo {year} {1997})}\BibitemShut {NoStop}%
\bibitem [{\citenamefont {Goldman}\ and\ \citenamefont {Nussinov}(1989)}]{Goldman:1989nd}%
  \BibitemOpen
  \bibfield  {author} {\bibinfo {author} {\bibfnamefont {I.}~\bibnamefont {Goldman}}\ and\ \bibinfo {author} {\bibfnamefont {S.}~\bibnamefont {Nussinov}},\ }\href {\doibase 10.1103/PhysRevD.40.3221} {\bibfield  {journal} {\bibinfo  {journal} {Phys. Rev. D}\ }\textbf {\bibinfo {volume} {40}},\ \bibinfo {pages} {3221} (\bibinfo {year} {1989})}\BibitemShut {NoStop}%
\bibitem [{\citenamefont {Bertone}\ and\ \citenamefont {Fairbairn}(2008)}]{Bertone:2007ae}%
  \BibitemOpen
  \bibfield  {author} {\bibinfo {author} {\bibfnamefont {G.}~\bibnamefont {Bertone}}\ and\ \bibinfo {author} {\bibfnamefont {M.}~\bibnamefont {Fairbairn}},\ }\href {\doibase 10.1103/PhysRevD.77.043515} {\bibfield  {journal} {\bibinfo  {journal} {Phys. Rev. D}\ }\textbf {\bibinfo {volume} {77}},\ \bibinfo {pages} {043515} (\bibinfo {year} {2008})},\ \Eprint {http://arxiv.org/abs/0709.1485} {arXiv:0709.1485 [astro-ph]} \BibitemShut {NoStop}%
\bibitem [{\citenamefont {Ciarcelluti}\ and\ \citenamefont {Sandin}(2011)}]{Ciarcelluti:2010ji}%
  \BibitemOpen
  \bibfield  {author} {\bibinfo {author} {\bibfnamefont {P.}~\bibnamefont {Ciarcelluti}}\ and\ \bibinfo {author} {\bibfnamefont {F.}~\bibnamefont {Sandin}},\ }\href {\doibase 10.1016/j.physletb.2010.11.021} {\bibfield  {journal} {\bibinfo  {journal} {Phys. Lett. B}\ }\textbf {\bibinfo {volume} {695}},\ \bibinfo {pages} {19} (\bibinfo {year} {2011})},\ \Eprint {http://arxiv.org/abs/1005.0857} {arXiv:1005.0857 [astro-ph.HE]} \BibitemShut {NoStop}%
\bibitem [{\citenamefont {Li}\ \emph {et~al.}(2012)\citenamefont {Li}, \citenamefont {Wang},\ and\ \citenamefont {Cheng}}]{Li:2012qf}%
  \BibitemOpen
  \bibfield  {author} {\bibinfo {author} {\bibfnamefont {X.}~\bibnamefont {Li}}, \bibinfo {author} {\bibfnamefont {F.}~\bibnamefont {Wang}}, \ and\ \bibinfo {author} {\bibfnamefont {K.~S.}\ \bibnamefont {Cheng}},\ }\href {\doibase 10.1088/1475-7516/2012/10/031} {\bibfield  {journal} {\bibinfo  {journal} {JCAP}\ }\textbf {\bibinfo {volume} {10}},\ \bibinfo {pages} {031} (\bibinfo {year} {2012})},\ \Eprint {http://arxiv.org/abs/1210.1748} {arXiv:1210.1748 [astro-ph.CO]} \BibitemShut {NoStop}%
\bibitem [{\citenamefont {Ellis}\ \emph {et~al.}(2018)\citenamefont {Ellis}, \citenamefont {H\"utsi}, \citenamefont {Kannike}, \citenamefont {Marzola}, \citenamefont {Raidal},\ and\ \citenamefont {Vaskonen}}]{Ellis:2018bkr}%
  \BibitemOpen
  \bibfield  {author} {\bibinfo {author} {\bibfnamefont {J.}~\bibnamefont {Ellis}}, \bibinfo {author} {\bibfnamefont {G.}~\bibnamefont {H\"utsi}}, \bibinfo {author} {\bibfnamefont {K.}~\bibnamefont {Kannike}}, \bibinfo {author} {\bibfnamefont {L.}~\bibnamefont {Marzola}}, \bibinfo {author} {\bibfnamefont {M.}~\bibnamefont {Raidal}}, \ and\ \bibinfo {author} {\bibfnamefont {V.}~\bibnamefont {Vaskonen}},\ }\href {\doibase 10.1103/PhysRevD.97.123007} {\bibfield  {journal} {\bibinfo  {journal} {Phys. Rev. D}\ }\textbf {\bibinfo {volume} {97}},\ \bibinfo {pages} {123007} (\bibinfo {year} {2018})},\ \Eprint {http://arxiv.org/abs/1804.01418} {arXiv:1804.01418 [astro-ph.CO]} \BibitemShut {NoStop}%
\bibitem [{\citenamefont {Nelson}\ \emph {et~al.}(2019)\citenamefont {Nelson}, \citenamefont {Reddy},\ and\ \citenamefont {Zhou}}]{Nelson:2018xtr}%
  \BibitemOpen
  \bibfield  {author} {\bibinfo {author} {\bibfnamefont {A.}~\bibnamefont {Nelson}}, \bibinfo {author} {\bibfnamefont {S.}~\bibnamefont {Reddy}}, \ and\ \bibinfo {author} {\bibfnamefont {D.}~\bibnamefont {Zhou}},\ }\href {\doibase 10.1088/1475-7516/2019/07/012} {\bibfield  {journal} {\bibinfo  {journal} {JCAP}\ }\textbf {\bibinfo {volume} {07}},\ \bibinfo {pages} {012} (\bibinfo {year} {2019})},\ \Eprint {http://arxiv.org/abs/1803.03266} {arXiv:1803.03266 [hep-ph]} \BibitemShut {NoStop}%
\bibitem [{\citenamefont {Ivanytskyi}\ \emph {et~al.}(2020)\citenamefont {Ivanytskyi}, \citenamefont {Sagun},\ and\ \citenamefont {Lopes}}]{PhysRevD.102.063028}%
  \BibitemOpen
  \bibfield  {author} {\bibinfo {author} {\bibfnamefont {O.}~\bibnamefont {Ivanytskyi}}, \bibinfo {author} {\bibfnamefont {V.}~\bibnamefont {Sagun}}, \ and\ \bibinfo {author} {\bibfnamefont {I.}~\bibnamefont {Lopes}},\ }\href {\doibase 10.1103/PhysRevD.102.063028} {\bibfield  {journal} {\bibinfo  {journal} {Phys. Rev. D}\ }\textbf {\bibinfo {volume} {102}},\ \bibinfo {pages} {063028} (\bibinfo {year} {2020})}\BibitemShut {NoStop}%
\bibitem [{\citenamefont {Das}\ \emph {et~al.}(2020)\citenamefont {Das}, \citenamefont {Kumar}, \citenamefont {Kumar}, \citenamefont {Kumar~Biswal}, \citenamefont {Nakatsukasa}, \citenamefont {Li},\ and\ \citenamefont {Patra}}]{Das:2020vng}%
  \BibitemOpen
  \bibfield  {author} {\bibinfo {author} {\bibfnamefont {H.~C.}\ \bibnamefont {Das}}, \bibinfo {author} {\bibfnamefont {A.}~\bibnamefont {Kumar}}, \bibinfo {author} {\bibfnamefont {B.}~\bibnamefont {Kumar}}, \bibinfo {author} {\bibfnamefont {S.}~\bibnamefont {Kumar~Biswal}}, \bibinfo {author} {\bibfnamefont {T.}~\bibnamefont {Nakatsukasa}}, \bibinfo {author} {\bibfnamefont {A.}~\bibnamefont {Li}}, \ and\ \bibinfo {author} {\bibfnamefont {S.~K.}\ \bibnamefont {Patra}},\ }\href {\doibase 10.1093/mnras/staa1435} {\bibfield  {journal} {\bibinfo  {journal} {Mon. Not. Roy. Astron. Soc.}\ }\textbf {\bibinfo {volume} {495}},\ \bibinfo {pages} {4893} (\bibinfo {year} {2020})},\ \Eprint {http://arxiv.org/abs/2002.00594} {arXiv:2002.00594 [nucl-th]} \BibitemShut {NoStop}%
\bibitem [{\citenamefont {Berezhiani}\ \emph {et~al.}(2021)\citenamefont {Berezhiani}, \citenamefont {Biondi}, \citenamefont {Mannarelli},\ and\ \citenamefont {Tonelli}}]{Berezhiani:2020zck}%
  \BibitemOpen
  \bibfield  {author} {\bibinfo {author} {\bibfnamefont {Z.}~\bibnamefont {Berezhiani}}, \bibinfo {author} {\bibfnamefont {R.}~\bibnamefont {Biondi}}, \bibinfo {author} {\bibfnamefont {M.}~\bibnamefont {Mannarelli}}, \ and\ \bibinfo {author} {\bibfnamefont {F.}~\bibnamefont {Tonelli}},\ }\href {\doibase 10.1140/epjc/s10052-021-09806-1} {\bibfield  {journal} {\bibinfo  {journal} {Eur. Phys. J. C}\ }\textbf {\bibinfo {volume} {81}},\ \bibinfo {pages} {1036} (\bibinfo {year} {2021})},\ \Eprint {http://arxiv.org/abs/2012.15233} {arXiv:2012.15233 [astro-ph.HE]} \BibitemShut {NoStop}%
\bibitem [{\citenamefont {Giangrandi}\ \emph {et~al.}(2023)\citenamefont {Giangrandi}, \citenamefont {Sagun}, \citenamefont {Ivanytskyi}, \citenamefont {Provid\^encia},\ and\ \citenamefont {Dietrich}}]{Giangrandi:2022wht}%
  \BibitemOpen
  \bibfield  {author} {\bibinfo {author} {\bibfnamefont {E.}~\bibnamefont {Giangrandi}}, \bibinfo {author} {\bibfnamefont {V.}~\bibnamefont {Sagun}}, \bibinfo {author} {\bibfnamefont {O.}~\bibnamefont {Ivanytskyi}}, \bibinfo {author} {\bibfnamefont {C.}~\bibnamefont {Provid\^encia}}, \ and\ \bibinfo {author} {\bibfnamefont {T.}~\bibnamefont {Dietrich}},\ }\href {\doibase 10.3847/1538-4357/ace104} {\bibfield  {journal} {\bibinfo  {journal} {Astrophys. J.}\ }\textbf {\bibinfo {volume} {953}},\ \bibinfo {pages} {115} (\bibinfo {year} {2023})},\ \Eprint {http://arxiv.org/abs/2209.10905} {arXiv:2209.10905 [astro-ph.HE]} \BibitemShut {NoStop}%
\bibitem [{\citenamefont {Kumar}\ \emph {et~al.}(2022)\citenamefont {Kumar}, \citenamefont {Das},\ and\ \citenamefont {Patra}}]{Kumar:2022amh}%
  \BibitemOpen
  \bibfield  {author} {\bibinfo {author} {\bibfnamefont {A.}~\bibnamefont {Kumar}}, \bibinfo {author} {\bibfnamefont {H.~C.}\ \bibnamefont {Das}}, \ and\ \bibinfo {author} {\bibfnamefont {S.~K.}\ \bibnamefont {Patra}},\ }\href {\doibase 10.1093/mnras/stac1013} {\bibfield  {journal} {\bibinfo  {journal} {Mon. Not. Roy. Astron. Soc.}\ }\textbf {\bibinfo {volume} {513}},\ \bibinfo {pages} {1820} (\bibinfo {year} {2022})},\ \Eprint {http://arxiv.org/abs/2203.02132} {arXiv:2203.02132 [astro-ph.HE]} \BibitemShut {NoStop}%
\bibitem [{\citenamefont {Shakeri}\ and\ \citenamefont {Karkevandi}(2024)}]{PhysRevD.109.043029}%
  \BibitemOpen
  \bibfield  {author} {\bibinfo {author} {\bibfnamefont {S.}~\bibnamefont {Shakeri}}\ and\ \bibinfo {author} {\bibfnamefont {D.~R.}\ \bibnamefont {Karkevandi}},\ }\href {\doibase 10.1103/PhysRevD.109.043029} {\bibfield  {journal} {\bibinfo  {journal} {Phys. Rev. D}\ }\textbf {\bibinfo {volume} {109}},\ \bibinfo {pages} {043029} (\bibinfo {year} {2024})}\BibitemShut {NoStop}%
\bibitem [{\citenamefont {Thakur}\ \emph {et~al.}(2024)\citenamefont {Thakur}, \citenamefont {Malik}, \citenamefont {Das}, \citenamefont {Jha},\ and\ \citenamefont {Provid\^encia}}]{Thakur:2023aqm}%
  \BibitemOpen
  \bibfield  {author} {\bibinfo {author} {\bibfnamefont {P.}~\bibnamefont {Thakur}}, \bibinfo {author} {\bibfnamefont {T.}~\bibnamefont {Malik}}, \bibinfo {author} {\bibfnamefont {A.}~\bibnamefont {Das}}, \bibinfo {author} {\bibfnamefont {T.~K.}\ \bibnamefont {Jha}}, \ and\ \bibinfo {author} {\bibfnamefont {C.}~\bibnamefont {Provid\^encia}},\ }\href {\doibase 10.1103/PhysRevD.109.043030} {\bibfield  {journal} {\bibinfo  {journal} {Phys. Rev. D}\ }\textbf {\bibinfo {volume} {109}},\ \bibinfo {pages} {043030} (\bibinfo {year} {2024})},\ \Eprint {http://arxiv.org/abs/2308.00650} {arXiv:2308.00650 [hep-ph]} \BibitemShut {NoStop}%
\bibitem [{\citenamefont {Anderson}\ \emph {et~al.}(2000)\citenamefont {Anderson}, \citenamefont {Haljan}, \citenamefont {Wieman},\ and\ \citenamefont {Cornell}}]{Anderson_2000}%
  \BibitemOpen
  \bibfield  {author} {\bibinfo {author} {\bibfnamefont {B.~P.}\ \bibnamefont {Anderson}}, \bibinfo {author} {\bibfnamefont {P.~C.}\ \bibnamefont {Haljan}}, \bibinfo {author} {\bibfnamefont {C.~E.}\ \bibnamefont {Wieman}}, \ and\ \bibinfo {author} {\bibfnamefont {E.~A.}\ \bibnamefont {Cornell}},\ }\href {\doibase 10.1103/physrevlett.85.2857} {\bibfield  {journal} {\bibinfo  {journal} {Physical Review Letters}\ }\textbf {\bibinfo {volume} {85}},\ \bibinfo {pages} {2857–2860} (\bibinfo {year} {2000})}\BibitemShut {NoStop}%
\bibitem [{\citenamefont {Dolan}\ and\ \citenamefont {Jackiw}(1974)}]{Dolan:1973qd}%
  \BibitemOpen
  \bibfield  {author} {\bibinfo {author} {\bibfnamefont {L.}~\bibnamefont {Dolan}}\ and\ \bibinfo {author} {\bibfnamefont {R.}~\bibnamefont {Jackiw}},\ }\href {\doibase 10.1103/PhysRevD.9.3320} {\bibfield  {journal} {\bibinfo  {journal} {Phys. Rev. D}\ }\textbf {\bibinfo {volume} {9}},\ \bibinfo {pages} {3320} (\bibinfo {year} {1974})}\BibitemShut {NoStop}%
\bibitem [{\citenamefont {Stoof}(1992)}]{Stoof:1992zz}%
  \BibitemOpen
  \bibfield  {author} {\bibinfo {author} {\bibfnamefont {H.~T.~C.}\ \bibnamefont {Stoof}},\ }\href {\doibase 10.1103/PhysRevA.45.8398} {\bibfield  {journal} {\bibinfo  {journal} {Phys. Rev. A}\ }\textbf {\bibinfo {volume} {45}},\ \bibinfo {pages} {8398} (\bibinfo {year} {1992})}\BibitemShut {NoStop}%
\bibitem [{\citenamefont {{Gr{\"u}ter}}\ \emph {et~al.}(1997)\citenamefont {{Gr{\"u}ter}}, \citenamefont {{Ceperley}},\ and\ \citenamefont {{Lalo{\"e}}}}]{1997PhRvL..79.3549G}%
  \BibitemOpen
  \bibfield  {author} {\bibinfo {author} {\bibfnamefont {P.}~\bibnamefont {{Gr{\"u}ter}}}, \bibinfo {author} {\bibfnamefont {D.}~\bibnamefont {{Ceperley}}}, \ and\ \bibinfo {author} {\bibfnamefont {F.}~\bibnamefont {{Lalo{\"e}}}},\ }\href {\doibase 10.1103/PhysRevLett.79.3549} {\bibfield  {journal} {\bibinfo  {journal} {Phys. Rev. Lett.}\ }\textbf {\bibinfo {volume} {79}},\ \bibinfo {pages} {3549} (\bibinfo {year} {1997})},\ \Eprint {http://arxiv.org/abs/cond-mat/9707028} {arXiv:cond-mat/9707028 [cond-mat.stat-mech]} \BibitemShut {NoStop}%
\bibitem [{\citenamefont {Jeon}\ and\ \citenamefont {Yaffe}(1996)}]{PhysRevD.53.5799}%
  \BibitemOpen
  \bibfield  {author} {\bibinfo {author} {\bibfnamefont {S.}~\bibnamefont {Jeon}}\ and\ \bibinfo {author} {\bibfnamefont {L.~G.}\ \bibnamefont {Yaffe}},\ }\href {\doibase 10.1103/PhysRevD.53.5799} {\bibfield  {journal} {\bibinfo  {journal} {Phys. Rev. D}\ }\textbf {\bibinfo {volume} {53}},\ \bibinfo {pages} {5799} (\bibinfo {year} {1996})}\BibitemShut {NoStop}%
\bibitem [{\citenamefont {Spergel}\ and\ \citenamefont {Steinhardt}(2000)}]{Spergel:1999mh}%
  \BibitemOpen
  \bibfield  {author} {\bibinfo {author} {\bibfnamefont {D.~N.}\ \bibnamefont {Spergel}}\ and\ \bibinfo {author} {\bibfnamefont {P.~J.}\ \bibnamefont {Steinhardt}},\ }\href {\doibase 10.1103/PhysRevLett.84.3760} {\bibfield  {journal} {\bibinfo  {journal} {Phys. Rev. Lett.}\ }\textbf {\bibinfo {volume} {84}},\ \bibinfo {pages} {3760} (\bibinfo {year} {2000})},\ \Eprint {http://arxiv.org/abs/astro-ph/9909386} {arXiv:astro-ph/9909386} \BibitemShut {NoStop}%
\bibitem [{\citenamefont {Tulin}\ and\ \citenamefont {Yu}(2018)}]{Tulin:2017ara}%
  \BibitemOpen
  \bibfield  {author} {\bibinfo {author} {\bibfnamefont {S.}~\bibnamefont {Tulin}}\ and\ \bibinfo {author} {\bibfnamefont {H.-B.}\ \bibnamefont {Yu}},\ }\href {\doibase 10.1016/j.physrep.2017.11.004} {\bibfield  {journal} {\bibinfo  {journal} {Phys. Rept.}\ }\textbf {\bibinfo {volume} {730}},\ \bibinfo {pages} {1} (\bibinfo {year} {2018})},\ \Eprint {http://arxiv.org/abs/1705.02358} {arXiv:1705.02358 [hep-ph]} \BibitemShut {NoStop}%
\bibitem [{\citenamefont {Ianni}\ \emph {et~al.}(2022)\citenamefont {Ianni}, \citenamefont {Mannarelli},\ and\ \citenamefont {Rossi}}]{Ianni:2021ynp}%
  \BibitemOpen
  \bibfield  {author} {\bibinfo {author} {\bibfnamefont {A.}~\bibnamefont {Ianni}}, \bibinfo {author} {\bibfnamefont {M.}~\bibnamefont {Mannarelli}}, \ and\ \bibinfo {author} {\bibfnamefont {N.}~\bibnamefont {Rossi}},\ }\href {\doibase 10.1016/j.rinp.2022.105544} {\bibfield  {journal} {\bibinfo  {journal} {Results Phys.}\ }\textbf {\bibinfo {volume} {38}},\ \bibinfo {pages} {105544} (\bibinfo {year} {2022})},\ \Eprint {http://arxiv.org/abs/2112.03755} {arXiv:2112.03755 [hep-ph]} \BibitemShut {NoStop}%
\bibitem [{\citenamefont {Nesti}\ \emph {et~al.}(2023)\citenamefont {Nesti}, \citenamefont {Salucci},\ and\ \citenamefont {Turini}}]{Nesti:2023tid}%
  \BibitemOpen
  \bibfield  {author} {\bibinfo {author} {\bibfnamefont {F.}~\bibnamefont {Nesti}}, \bibinfo {author} {\bibfnamefont {P.}~\bibnamefont {Salucci}}, \ and\ \bibinfo {author} {\bibfnamefont {N.}~\bibnamefont {Turini}},\ }\href {\doibase 10.3390/astronomy2020007} {\bibfield  {journal} {\bibinfo  {journal} {Astronomy}\ }\textbf {\bibinfo {volume} {2}},\ \bibinfo {pages} {90} (\bibinfo {year} {2023})},\ \Eprint {http://arxiv.org/abs/2308.02004} {arXiv:2308.02004 [hep-ph]} \BibitemShut {NoStop}%
\bibitem [{\citenamefont {Ferioli}\ \emph {et~al.}(2019)\citenamefont {Ferioli}, \citenamefont {Semeghini}, \citenamefont {Masi}, \citenamefont {Giusti}, \citenamefont {Modugno}, \citenamefont {Inguscio}, \citenamefont {Gallem\'{\i}}, \citenamefont {Recati},\ and\ \citenamefont {Fattori}}]{PhysRevLett.122.090401}%
  \BibitemOpen
  \bibfield  {author} {\bibinfo {author} {\bibfnamefont {G.}~\bibnamefont {Ferioli}}, \bibinfo {author} {\bibfnamefont {G.}~\bibnamefont {Semeghini}}, \bibinfo {author} {\bibfnamefont {L.}~\bibnamefont {Masi}}, \bibinfo {author} {\bibfnamefont {G.}~\bibnamefont {Giusti}}, \bibinfo {author} {\bibfnamefont {G.}~\bibnamefont {Modugno}}, \bibinfo {author} {\bibfnamefont {M.}~\bibnamefont {Inguscio}}, \bibinfo {author} {\bibfnamefont {A.}~\bibnamefont {Gallem\'{\i}}}, \bibinfo {author} {\bibfnamefont {A.}~\bibnamefont {Recati}}, \ and\ \bibinfo {author} {\bibfnamefont {M.}~\bibnamefont {Fattori}},\ }\href {\doibase 10.1103/PhysRevLett.122.090401} {\bibfield  {journal} {\bibinfo  {journal} {Phys. Rev. Lett.}\ }\textbf {\bibinfo {volume} {122}},\ \bibinfo {pages} {090401} (\bibinfo {year} {2019})}\BibitemShut {NoStop}%
\bibitem [{\citenamefont {Norcia}\ \emph {et~al.}(2021)\citenamefont {Norcia}, \citenamefont {Politi}, \citenamefont {Klaus}, \citenamefont {Poli}, \citenamefont {Sohmen}, \citenamefont {Mark}, \citenamefont {Bisset}, \citenamefont {Santos},\ and\ \citenamefont {Ferlaino}}]{norcia2021tds}%
  \BibitemOpen
  \bibfield  {author} {\bibinfo {author} {\bibfnamefont {M.~A.}\ \bibnamefont {Norcia}}, \bibinfo {author} {\bibfnamefont {C.}~\bibnamefont {Politi}}, \bibinfo {author} {\bibfnamefont {L.}~\bibnamefont {Klaus}}, \bibinfo {author} {\bibfnamefont {E.}~\bibnamefont {Poli}}, \bibinfo {author} {\bibfnamefont {M.}~\bibnamefont {Sohmen}}, \bibinfo {author} {\bibfnamefont {M.~J.}\ \bibnamefont {Mark}}, \bibinfo {author} {\bibfnamefont {R.~N.}\ \bibnamefont {Bisset}}, \bibinfo {author} {\bibfnamefont {L.}~\bibnamefont {Santos}}, \ and\ \bibinfo {author} {\bibfnamefont {F.}~\bibnamefont {Ferlaino}},\ }\href {https://doi.org/10.1038/s41586-021-03725-7} {\bibfield  {journal} {\bibinfo  {journal} {Nature}\ }\textbf {\bibinfo {volume} {596}},\ \bibinfo {pages} {357} (\bibinfo {year} {2021})}\BibitemShut {NoStop}%
\bibitem [{\citenamefont {Bland}\ \emph {et~al.}(2022)\citenamefont {Bland}, \citenamefont {Poli}, \citenamefont {Politi}, \citenamefont {Klaus}, \citenamefont {Norcia}, \citenamefont {Ferlaino}, \citenamefont {Santos},\ and\ \citenamefont {Bisset}}]{Bland2022tds}%
  \BibitemOpen
  \bibfield  {author} {\bibinfo {author} {\bibfnamefont {T.}~\bibnamefont {Bland}}, \bibinfo {author} {\bibfnamefont {E.}~\bibnamefont {Poli}}, \bibinfo {author} {\bibfnamefont {C.}~\bibnamefont {Politi}}, \bibinfo {author} {\bibfnamefont {L.}~\bibnamefont {Klaus}}, \bibinfo {author} {\bibfnamefont {M.}~\bibnamefont {Norcia}}, \bibinfo {author} {\bibfnamefont {F.}~\bibnamefont {Ferlaino}}, \bibinfo {author} {\bibfnamefont {L.}~\bibnamefont {Santos}}, \ and\ \bibinfo {author} {\bibfnamefont {R.}~\bibnamefont {Bisset}},\ }\href {https://doi.org/10.1103/PhysRevLett.128.195302} {\bibfield  {journal} {\bibinfo  {journal} {Physical Review Letters}\ }\textbf {\bibinfo {volume} {128}},\ \bibinfo {pages} {195302} (\bibinfo {year} {2022})}\BibitemShut {NoStop}%
\bibitem [{\citenamefont {Norcia}\ \emph {et~al.}(2022)\citenamefont {Norcia}, \citenamefont {Poli}, \citenamefont {Politi}, \citenamefont {Klaus}, \citenamefont {Bland}, \citenamefont {Mark}, \citenamefont {Santos}, \citenamefont {Bisset},\ and\ \citenamefont {Ferlaino}}]{norcia2022cao}%
  \BibitemOpen
  \bibfield  {author} {\bibinfo {author} {\bibfnamefont {M.~A.}\ \bibnamefont {Norcia}}, \bibinfo {author} {\bibfnamefont {E.}~\bibnamefont {Poli}}, \bibinfo {author} {\bibfnamefont {C.}~\bibnamefont {Politi}}, \bibinfo {author} {\bibfnamefont {L.}~\bibnamefont {Klaus}}, \bibinfo {author} {\bibfnamefont {T.}~\bibnamefont {Bland}}, \bibinfo {author} {\bibfnamefont {M.~J.}\ \bibnamefont {Mark}}, \bibinfo {author} {\bibfnamefont {L.}~\bibnamefont {Santos}}, \bibinfo {author} {\bibfnamefont {R.~N.}\ \bibnamefont {Bisset}}, \ and\ \bibinfo {author} {\bibfnamefont {F.}~\bibnamefont {Ferlaino}},\ }\href {\doibase 10.1103/PhysRevLett.129.040403} {\bibfield  {journal} {\bibinfo  {journal} {Phys. Rev. Lett.}\ }\textbf {\bibinfo {volume} {129}},\ \bibinfo {pages} {040403} (\bibinfo {year} {2022})}\BibitemShut {NoStop}%
\bibitem [{\citenamefont {Klaus}\ \emph {et~al.}(2022)\citenamefont {Klaus}, \citenamefont {Bland}, \citenamefont {Poli}, \citenamefont {Politi}, \citenamefont {Lamporesi}, \citenamefont {Casotti}, \citenamefont {Bisset}, \citenamefont {Mark},\ and\ \citenamefont {Ferlaino}}]{klaus2022oov}%
  \BibitemOpen
  \bibfield  {author} {\bibinfo {author} {\bibfnamefont {L.}~\bibnamefont {Klaus}}, \bibinfo {author} {\bibfnamefont {T.}~\bibnamefont {Bland}}, \bibinfo {author} {\bibfnamefont {E.}~\bibnamefont {Poli}}, \bibinfo {author} {\bibfnamefont {C.}~\bibnamefont {Politi}}, \bibinfo {author} {\bibfnamefont {G.}~\bibnamefont {Lamporesi}}, \bibinfo {author} {\bibfnamefont {E.}~\bibnamefont {Casotti}}, \bibinfo {author} {\bibfnamefont {R.~N.}\ \bibnamefont {Bisset}}, \bibinfo {author} {\bibfnamefont {M.~J.}\ \bibnamefont {Mark}}, \ and\ \bibinfo {author} {\bibfnamefont {F.}~\bibnamefont {Ferlaino}},\ }\href {https://doi.org/10.1038/s41567-022-01793-8} {\bibfield  {journal} {\bibinfo  {journal} {Nature Physics}\ }\textbf {\bibinfo {volume} {18}},\ \bibinfo {pages} {1453} (\bibinfo {year} {2022})}\BibitemShut {NoStop}%
\bibitem [{\citenamefont {Poli}\ \emph {et~al.}(2023)\citenamefont {Poli}, \citenamefont {Bland}, \citenamefont {White}, \citenamefont {Mark}, \citenamefont {Ferlaino}, \citenamefont {Trabucco},\ and\ \citenamefont {Mannarelli}}]{Poli:2023vyp}%
  \BibitemOpen
  \bibfield  {author} {\bibinfo {author} {\bibfnamefont {E.}~\bibnamefont {Poli}}, \bibinfo {author} {\bibfnamefont {T.}~\bibnamefont {Bland}}, \bibinfo {author} {\bibfnamefont {S.~J.~M.}\ \bibnamefont {White}}, \bibinfo {author} {\bibfnamefont {M.~J.}\ \bibnamefont {Mark}}, \bibinfo {author} {\bibfnamefont {F.}~\bibnamefont {Ferlaino}}, \bibinfo {author} {\bibfnamefont {S.}~\bibnamefont {Trabucco}}, \ and\ \bibinfo {author} {\bibfnamefont {M.}~\bibnamefont {Mannarelli}},\ }\href {\doibase 10.1103/PhysRevLett.131.223401} {\bibfield  {journal} {\bibinfo  {journal} {Phys. Rev. Lett.}\ }\textbf {\bibinfo {volume} {131}},\ \bibinfo {pages} {223401} (\bibinfo {year} {2023})},\ \Eprint {http://arxiv.org/abs/2306.09698} {arXiv:2306.09698 [cond-mat.quant-gas]} \BibitemShut {NoStop}%
\end{thebibliography}
%

%
\end{document}